\documentclass[twocolumn]{aastex702}
\graphicspath{{figs/}}

\defcitealias{panuzzo24}{Gaia Collaboration et al.}

\newcommand{\Mdw}{\dot M_{\rm w}}
\newcommand{\fEdd}{f_{\rm Edd}}
\newcommand{\Msunyr}{M_\odot\,{\rm yr^{-1}}}
\newcommand{\kms}{km\,s$^{-1}$}

\shorttitle{Emission from Wind Accretion onto Gaia BH3}
\shortauthors{Cappelluti \& Pacucci}

\begin{document}
\title{Optical and Ultraviolet Emission from the Gas Photoionized by Gaia BH3: \\Archival Constraints and Predictions for the 2029 Periastron}

\author[orcid=0000-0002-1697-186X]{Nico Cappelluti}
\affiliation{Department of Physics, University of Miami, Coral Gables, FL 33124, USA}
\email[show]{ncappelluti@miami.edu}
\correspondingauthor{N. Cappelluti}

\author[orcid=0000-0001-9879-7780]{Fabio Pacucci}
\affiliation{Center for Astrophysics $\vert$ Harvard \& Smithsonian, 60 Garden St, Cambridge, MA 02138, USA}
\email[show]{fabio.pacucci@cfa.harvard.edu}

\begin{abstract}
Gaia~BH3, a $32.7\,M_\odot$ black hole (BH) on an 11.6-yr, $e=0.73$ orbit around a metal-poor giant at $\approx590$~pc, accretes from its wind. Neither its mass-loss rate $\Mdw$ nor the fraction of captured gas reaching the BH is known. We ask whether optical and ultraviolet lines from gas photoionized by the hot flow are observable and what they reveal. Unlike X-ray and radio data, they count the ionizing photons the gas absorbs and trace its density and velocity. With \textsc{Cloudy} we compute the H$\alpha$ and He\,II~4686 emission of the wind and inflow along the orbit for $\Mdw=10^{-11}$--$10^{-8}\,\Msunyr$, Eddington ratios $\fEdd=10^{-8}$--$10^{-5}$ and wind speeds of 30 (slow) and 100~\kms\ (fast), and test each model against archival UVES and ESPRESSO spectra. $\fEdd$ mainly sets the lines; in dense models they come from gas falling freely onto the BH, spread over thousands of \kms. The archival spectra exclude $\Mdw>2\times10^{-9}\,\Msunyr$ at $\fEdd=2\times10^{-6}$ and $>2.6\times10^{-10}\,\Msunyr$ at $10^{-5}$ (95\%, slow wind), Eddington ratios already excluded by the apastron Chandra limit, $\fEdd<7.6\times10^{-7}$. Of two simulated campaigns, only 25 epochs around the 2029 periastron go beyond existing limits. For a slow wind, whose capture rate is constant, a $\ge5\sigma$ H$\alpha$ detection would imply an accretion state 2--3 times above the 2024 limit. For a fast wind, whose captured fraction rises 14-fold toward periastron, the campaign detects $\Mdw\simeq10^{-8}\,\Msunyr$ for $\fEdd\gtrsim1.2\times10^{-6}$, which the apastron limit allows. Below the Chandra limit, a slow wind produces no detectable line.
\end{abstract}


\section{Introduction}
Gaia~BH3 is the most massive stellar-mass black hole (BH) known in the Galaxy (\citetalias{panuzzo24} \citeyear{panuzzo24}), only 590~pc from Earth. It is a $32.7\,M_\odot$ dark companion to a very metal-poor ($[{\rm Fe/H}]=-2.56$) giant on an 11.6-yr orbit with $e=0.73$, with a separation that varies between 4.5 and 28.6~AU. The giant (with stellar parameters $M_\star=0.76\,M_\odot$, $R_\star=4.9\,R_\odot$, $T_{\rm eff}=5212$~K) loses mass through a cool wind, part of which the BH must capture. For a mass-loss rate in the range $\Mdw=10^{-11}$--$10^{-8}\,\Msunyr$, the Bondi--Hoyle--Lyttleton (BHL) capture rate is $\dot m_{\rm BHL}\simeq3\times10^{-6}$--$3\times10^{-3}$ in units of the Eddington accretion rate $\dot M_{\rm Edd}=L_{\rm Edd}/(0.1c^2)$ (see \S\ref{sec:model}). Even if all of this gas were accreted, the flow would be in the advection-dominated accretion flow regime \citep[ADAF;][]{narayan95}, whose radiation should photoionize the surrounding wind and inflow, producing recombination and collisionally excited lines. How much of the captured gas reaches the radiating flow, and how efficiently that flow radiates, remain unknown. For this reason, we keep two rates distinct: $\dot m_{\rm BHL}$, the capture rate set by the wind and the orbit, and $\fEdd\equiv\dot M/\dot M_{\rm Edd}$, the accretion rate that reaches the black hole, which sets the spectrum of the flow (\S\ref{sec:model}); their ratio $\fEdd/\dot m_{\rm BHL}$ is the fraction of the captured gas that is accreted. A 12.6~ks Chandra observation, executed on May 25, 2024 (i.e., with the BH at apastron) detected no X-rays, thus placing an upper limit of $L_{2-10}<2.1\times10^{29}$~erg\,s$^{-1}$ \citep{cappelluti24}. Assuming an ADAF spectrum (see \S\ref{sec:model}), this upper limit requires $\fEdd<7.6\times10^{-7}$, which is below $\dot m_{\rm BHL}$ for every $\Mdw$ in the range considered. However, this bound refers to a single epoch at the largest separation and provides no information about how much gas the giant loses. Optical emission lines from gas photoionized by the BH depend on both the wind mass-loss rate $\Mdw$ and the ionizing luminosity, and we can observe them at any orbital phase.

Gaia~BH3 is a clean test of accretion at very low rates. In X-ray binaries and active nuclei, we can measure the emission but not the gas supply; here, however, the BH mass, orbit, inclination, and distance are known, and the supply is the wind of a giant that we can study in its own spectrum. We can therefore compute the capture rate, and a detection of the accretion flow, or a limit on it, gives $\fEdd$ for the adopted spectrum and hence the fraction $\fEdd/\dot m_{\rm BHL}$ of the captured gas that is accreted. The X-ray limit already requires that less than $\approx 20$\% of the captured gas reaches the black hole if $\Mdw=10^{-11}\,\Msunyr$, and less than $2\times10^{-4}$ of it if $\Mdw=10^{-8}\,\Msunyr$. With $L_X/L_{\rm Edd}<5\times10^{-11}$, the system lies at least a factor of 60 below A0620-00, one of the faintest black-hole binaries detected in radio \citep{gallo06}, and approaches the regime of Sgr~A* and of the quiescent nuclei of most galaxies, where models of hot accretion flows \citep{yuan14} have extremely limited direct tests. Importantly, the eccentric orbit changes the separation by a factor of 6.4 on a known schedule. For a fast wind, this reduction in distance increases the captured fraction by more than tenfold toward periastron (\S\ref{sec:model}). Hence, we can measure the flow's response to a known change in its supply.

How efficiently captured gas is turned into radiation at these rates also sets the expected emission of the Galaxy's dormant black holes. The Milky Way should contain about $10^8$ stellar-mass black holes, most of them single \citep{Elbert_2018, olejak20}. Gaia has found a few in wide binaries \citep{elbadry23a,elbadry23b,panuzzo24}, and predictions for black holes accreting from their surroundings \citep[e.g.][]{agol02} depend on the efficiency we seek to estimate here. Mass loss on the red-giant branch at low metallicity is calibrated mostly on the integrated loss of globular-cluster stars \citep{mcdonald15}, whereas the line emission of the BH-illuminated wind, and the X-ray and radio emission of the flow it feeds, depend on the present mass-loss rate of a single metal-poor giant. Gaia~BH3 is also a nearby relative of the high-mass black holes detected in gravitational waves, which \citet{panuzzo24} suggest descend from metal-poor massive stars. The existing X-ray and radio limits already constrain the options: on the fundamental plane of black-hole activity \citep{merloni03}, the 15~GHz VLA upper limit \citep{sjouwerman24} lies a factor of 27 below the radio luminosity expected at the Chandra limit for a flat radio spectrum, so either $L_X\lesssim10^{27}$~erg\,s$^{-1}$ or Gaia~BH3 is radio-quiet compared with other quiescent black-hole binaries.

This study aims not to improve the X-ray and radio limits, but to test whether a different emission process is observable: the optical and ultraviolet lines of the gas that the BH photoionizes. The lines depend on quantities the other bands do not measure: (i) the number of ionizing photons the gas absorbs, and (ii) the density and velocity of the wind and inflow, i.e., the gas supply itself. We compute this emission along the orbit with a grid of photoionization models (\S\ref{sec:model}), compare it with archival UVES and ESPRESSO spectra (\S\ref{sec:data} and \S\ref{sec:constraints}), and predict what two spectroscopic campaigns can detect (\S\ref{sec:campaigns}): scenario~A, monthly monitoring in 2026 and 2027, and scenario~B, a dense campaign around the 2029 November periastron. Section~\ref{sec:discussion} discusses the results, including what the lines measure and how they relate to the X-ray and radio limits.

\section{Theoretical Model}\label{sec:model}
\subsection{Binary, wind and capture}
In our model, we adopt the \citetalias{panuzzo24} \citeyearpar{panuzzo24} orbit ($P=4253.1$~d, $e=0.7291$, $i=110.58^\circ$, $\omega=77.34^\circ$, $M_{\rm BH}=32.7\,M_\odot$, $M_\star=0.76\,M_\odot$), for which the separation $r$ ranges from 4.5~AU at periastron to $r_{\rm apo}=28.6$~AU.
Throughout, we describe the BH orbit relative to the giant; in the center-of-mass frame, the giant moves because the BH is 43 times more massive.
The giant's wind is spherical and accelerates according to the $\beta$ velocity law \citep{castor79,lamers99}, \begin{equation} v_{\rm w}(d)=v_0+(v_\infty-v_0)\left(1-\frac{R_\star}{d}\right)^{\beta_{\rm w}}, \end{equation} with $v_0=5$~\kms\ at the stellar surface and $\beta_{\rm w}=1$. Here $d$ is the distance from the giant, and mass conservation gives the wind density $\rho_{\rm w}=\Mdw/(4\pi d^2v_{\rm w})$. The BH never comes closer than $\simeq200\,R_\star$, where $v_{\rm w}\ge0.995\,v_\infty$, so only the terminal speed matters. The terminal speed of this giant's wind has not been measured, and it need not stay the same along the orbit. Our baseline calculation uses a slow wind, $v_\infty=30$~\kms, and we also use a fast wind, $ v_\infty=100$~\kms, to see how a faster wind affects the predictions. The \citet{reimers75} rate, $\dot M=4\times10^{-13}\,\eta\,(L/L_\odot)(R/R_\odot)(M/M_\odot)^{-1}\,\Msunyr$, gives $\Mdw\approx2\times10^{-11}\,\Msunyr$ for this star ($L\simeq16\,L_\odot$) with $\eta=0.5$, the value calibrated on globular-cluster giants \citep{mcdonald15}; direct measurements of winds from metal-poor giants exist only for stars one to two decades more luminous \citep{dupree09}, so we explore $\Mdw=10^{-11}$--$10^{-8}\,\Msunyr$.

\begin{figure*}[!tp]
\centering\includegraphics[width=\textwidth]{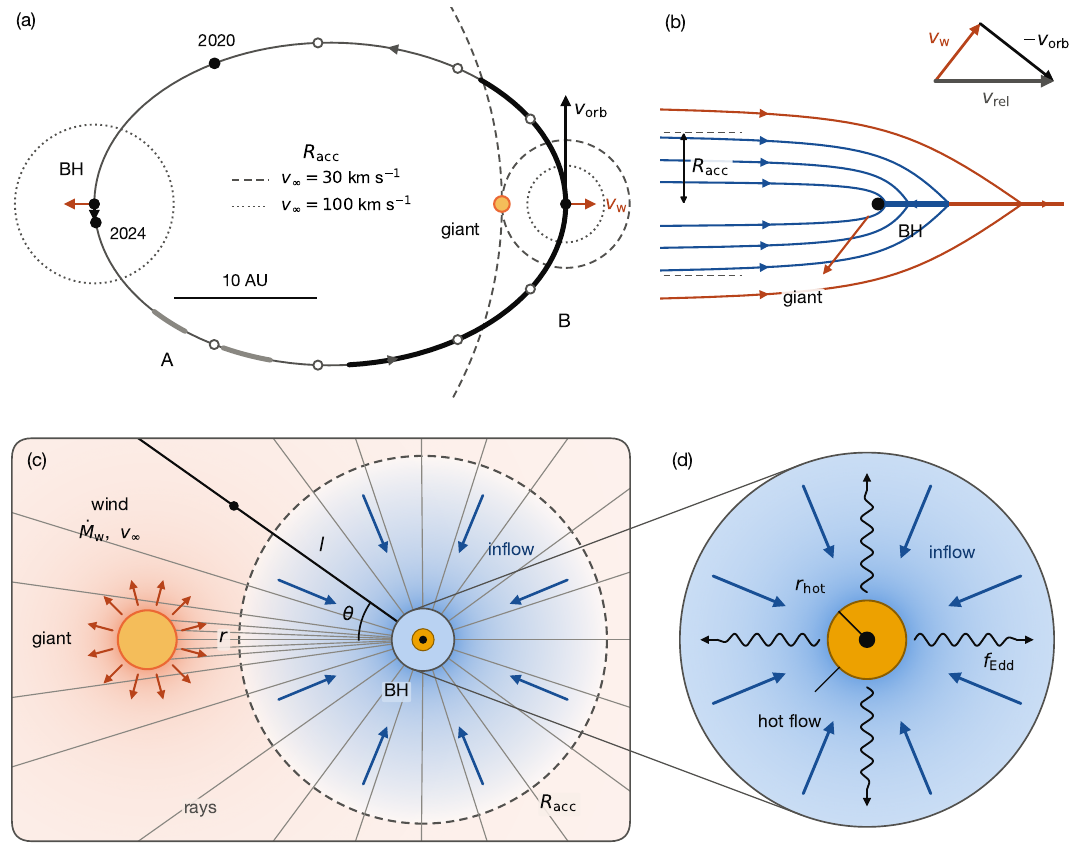}
\caption{Model geometry. (a) Orbit of the BH relative to the giant, to scale (BH and stellar sizes are not to scale). Open circles: the separations of the model grid. Filled circles: the BH at the archival-data epochs (2020, UVES; 2024, apastron, UVES, ESPRESSO, and Chandra). Thick arcs: scenario~A (gray; the 2026 and 2027 seasons) and scenario~B (black; 2028--2030, around the 2029 November periastron). Arrows at the apsides: the BH's orbital velocity relative to the giant, $v_{\rm orb}$, and the wind velocity at the BH, $v_{\rm w}$, on a common scale. Circles around the BH: the accretion radius $R_{\rm acc}$ for $v_\infty=30$~\kms\ (dashed; capped at the separation, \S\ref{sec:model}) and 100~\kms\ (dotted). (b) Wind capture in the BH frame, in units of $R_{\rm acc}$. The wind arrives with relative velocity $v_{\rm rel}$; streamlines passing within $R_{\rm acc}$ of the axis (blue) are accreted, while the others (red) escape. This picture shows only $R_{\rm acc}$ and the accretion rate; it does not model the wake. (c) The computed geometry (not to scale). The wind is spherical about the giant; inside $R_{\rm acc}$ (dashed) the gas falls freely onto the BH (blue arrows). We compute photoionization models along rays from the BH (gray), sampled finely near the BH--giant axis; one ray (black) shows the coordinates $l$ and $\theta$. Rays that meet the giant end at its surface. (d) The neighborhood of the BH. Inside $r_{\rm hot}$, the inflow is the hot accretion flow, which emits the ionizing spectrum set by $\fEdd$ (wavy arrows) and no lines. For scale, $R_\star=0.023$~AU, $r=4.5$--$28.6$~AU and $r_{\rm hot}=10^3\,R_g=3\times10^{-4}$~AU.}
\label{fig:cartoon}
\end{figure*}

In the frame of the BH, which moves relative to the giant with the orbital velocity $v_{\rm orb}=[GM_{\rm tot}(2/r-1/a)]^{1/2}=17$--107~\kms\ between apastron and periastron, the wind arrives with 
\begin{equation} 
v_{\rm rel}^2=v_{\rm w}^2+v_{\rm orb}^2+c_s^2 \, ,
\end{equation} 
where $c_s=12$~\kms\ is the isothermal sound speed of photoionized gas at $10^4$~K \citep{osterbrock06}. The sum in quadrature is exact at the apsides, where $v_{\rm rel}$ is inclined to the BH--giant direction by $\arctan(v_{\rm orb}/v_{\rm w})$, i.e., by $29^\circ$ at apastron and $74^\circ$ at periastron for the slow wind (Fig.~\ref{fig:cartoon}b); the thermal term contributes at most 11\% of $v_{\rm rel}^2$. The BH captures the wind within the accretion radius $R_{\rm acc}=2GM_{\rm BH}/v_{\rm rel}^2$ (Fig.~\ref{fig:cartoon}a,b) at the BHL rate 
\begin{equation} 
\dot M_{\rm BHL}=\pi R_{\rm acc}^2\rho_{\rm w}v_{\rm rel}=\frac{v_{\rm rel}}{4v_{\rm w}}\left(\frac{R_{\rm acc}}{r}\right)^2\Mdw \, . 
\end{equation} 
For the slow wind $2GM_{\rm BH}/v_{\rm rel}^2$ is 44~AU at apastron and 4.6~AU at periastron, larger than the separation at every phase: the wind is bound to the BH from the moment it leaves the giant, so that the BHL picture of a uniform flow past a point mass no longer applies. In this particular case, the BHL rate practically approaches $\Mdw$. This regime is typically described as wind Roche-lobe overflow \citep{mohamed07}: the slow wind progressively fills the giant's Roche lobe and is thus funneled toward the companion (via the inner Lagrangian point). Hydrodynamic simulations of slow winds in binaries find that a few tens of percent of the wind, and at most about half, reach the accretor, with the rest leaving the system \citep{abate13,saladino18}. We therefore set $R_{\rm acc}=r$, so the inflow region extends to the giant's location, and take the captured fraction to be 25\% of $\Mdw$ at every phase. The capture rate is then $\dot m_{\rm BHL}=3.4\times10^{-6}\,(\Mdw/10^{-11}\,\Msunyr)$ in Eddington units, independent of phase, and the orbit enters only through the density of the wind that the BH illuminates. For the fast wind neither cap applies: $R_{\rm acc}$ grows from $0.19\,r$ (5.6~AU) at apastron to $0.60\,r$ (2.7~AU) at periastron, and the captured fraction from 1\% to 13\% of $\Mdw$, a factor of 13.6. Inside $R_{\rm acc}$, the gas falls freely and spherically onto the BH. At each point along a ray, the density is the larger of the wind density and the free-fall inflow density; the inflow dominates within $0.07$--$0.56$ of the separation, depending on the phase, direction, and $v_\infty$. Inside $r_{\rm hot}=10^3\,R_g$ ($10^5\,R_g$ in a variant) the inflow is the hot flow itself and emits no lines (Fig.~\ref{fig:cartoon}c,d).

\subsection{Ionizing source}
We compute the BH spectrum with the public code LLAGNSED \citep{pesce21}, an analytic ADAF model that follows \citet{mahadevan97}. Its main physical parameters are the black hole mass, which we fix at Gaia BH3's known value, and the Eddington ratio $\fEdd\equiv\dot M(R_{\rm S})/\dot M_{\rm Edd}$, the accretion rate at one Schwarzschild radius $R_{\rm S}$ in units of $\dot M_{\rm Edd}=L_{\rm Edd}/(0.1c^2)$. The factor 0.1 only sets the unit of $\fEdd$, as in \citet{narayan95} and \citet{mahadevan97}; the flow itself is radiatively inefficient, and its radiative efficiency $\eta_{\rm rad}\equiv L_{\rm tot}/(\fEdd\dot M_{\rm Edd}c^2)$ rises from $4\times10^{-5}$ at $\fEdd=10^{-8}$ to $10^{-2}$ at $10^{-5}$ (Table~\ref{tab:sed}). Additional model parameters that can be tuned are the index of the radial accretion-rate profile, $s=0.5$ ($\dot M\propto r^{s}$), the viscosity parameter $\alpha=0.2$ \citep{shakura73}, the ratio of gas to magnetic pressure $\beta=10$, and the fraction of the viscous heating that goes into the electrons, $\delta=0.3$. 

The fraction of the viscous heating that is advected, $f=1-Q^-/Q^+$ with $Q^+$ the heating and $Q^-$ the radiative cooling rate, depends on the temperature and density of the flow, which depend on $f$; we use the self-consistent solution \citep{narayan95}, $f=0.95$--0.75 for $\fEdd=5\times10^{-7}$--$10^{-5}$, and $f=1$ at $\fEdd\le10^{-7}$, where the solution is $f\ge0.987$ (Table~\ref{tab:sed}). Over the seven Eddington ratios of the grid, $\fEdd=10^{-8}$--$10^{-5}$, the rate of hydrogen-ionizing photons $Q({\rm H})$ rises from $10^{35}$ to $7\times10^{42}$~s$^{-1}$ \citep{llerena25, pacucci25}. For these spectra the Chandra limit corresponds to $\fEdd<7.6\times10^{-7}$, between the grid values $7\times10^{-7}$ and $10^{-6}$. The fraction of captured gas that reaches the BH, $\epsilon\equiv\fEdd/\dot m_{\rm BHL}$, is unknown, so we keep $\fEdd$ and $\Mdw$ independent. Because $\dot M\propto r^{s}$, the hot flow takes in $(r_{\rm hot}/R_{\rm S})^{s}\approx22$ times more gas at $r_{\rm hot}=10^3\,R_g$ than reaches the BH, and loses the difference to an outflow. Together with the ADAF spectrum, $\epsilon$ sets the radiative efficiency of \S1, the luminosity per unit captured mass.

\begin{table}[!tbp]
\caption{The ADAF spectra of the grid: advected fraction $f$, hydrogen-ionizing photon rate, 2--10~keV and total luminosity}\label{tab:sed}
\centering\begin{tabular}{lcccc}
\hline
$f_{\rm Edd}$ & $f$ & $Q({\rm H})$ & $L_{2-10}$ & $L_{\rm tot}$\\
 & & [s$^{-1}$] & [erg\,s$^{-1}$] & [erg\,s$^{-1}$]\\
\hline
$10^{-8}$ & 1 & $1.0\times10^{35}$ & $3.4\times10^{25}$ & $1.6\times10^{28}$\\
$10^{-7}$ & 1 & $2.7\times10^{37}$ & $3.4\times10^{27}$ & $1.6\times10^{30}$\\
$5\times10^{-7}$ & 0.946 & $5.0\times10^{39}$ & $8.8\times10^{28}$ & $3.5\times10^{31}$\\
$7\times10^{-7}$ & 0.930 & $1.4\times10^{40}$ & $1.8\times10^{29}$ & $6.5\times10^{31}$\\
$10^{-6}$ & 0.910 & $3.9\times10^{40}$ & $3.7\times10^{29}$ & $1.2\times10^{32}$\\
$2\times10^{-6}$ & 0.864 & $2.4\times10^{41}$ & $1.6\times10^{30}$ & $3.9\times10^{32}$\\
$10^{-5}$ & 0.754 & $7.3\times10^{42}$ & $4.9\times10^{31}$ & $4.0\times10^{33}$\\
\hline
\end{tabular}
\end{table}

\subsection{Photoionization calculations and line profiles}
\textsc{Cloudy} is a one-dimensional code: it follows radiation from a central source outward along a single direction, with gas density given as a function of distance from the source. In this scenario, the source is the BH, while the density depends on the distance from the giant. The density along a direction from the BH depends on the angle $\theta$ between that direction and the axis between the BH and the giant (Fig.~\ref{fig:cartoon}c). This geometry describes an X-ray source embedded off-center in a stellar wind, first treated for massive X-ray binaries by \citet{hatchett77}. We therefore build each model from 17--19 one-dimensional \textsc{Cloudy} models (C23.01; \citealt{chatzikos23}), one per direction, each with the density along that ray tabulated as a function of the distance $l$ from the BH, and sum their line luminosities weighted by the solid angle that each ray represents. The directions are spaced finely near the BH--giant axis, where the wind density changes fastest. Rays that point at the giant end at its surface; others stop where the gas cools to 1500~K. Along each ray, the density is the larger of the wind and free-fall densities (\S\ref{sec:model}) and spans more than ten decades, from the cusp around the BH to the base of the wind. The rays are independent, with no diffuse radiation exchanged between them. The gas uses the \citet{grevesse10} solar mixture scaled to $[{\rm Fe/H}]=-2.56$ and $[\alpha/{\rm Fe}]=+0.4$, with no grains, and with the cosmic-ray background and the CMB included, and each ray is iterated to convergence.

The baseline grid has six separations ($r/r_{\rm apo}=1$, 0.79, 0.6, 0.35, 0.22 and 0.157; 0.79 is that of the 2020 UVES epoch), four values of $\Mdw$ and seven of $\fEdd$, i.e.\ 168 models. Three variant grids of 150 models each change $(v_\infty,r_{\rm hot})$ to (100~\kms, $10^3\,R_g$), (30~\kms, $10^5\,R_g$) and (100~\kms, $10^5\,R_g$), and 126 slab models describe a circularized cool ring at 0.1--2~AU with $n_{\rm H}=10^{8}$--$10^{12}$~cm$^{-3}$.

Every run completed and reproduced its input density law to within 0.013~dex; 14 slab runs required a smaller minimum zone thickness. Integrated over the rays, the H$\alpha$ luminosity is 0.66--1.0 of the analytic case-B value for $\fEdd\ge5\times10^{-7}$ and $\Mdw\le10^{-9}\,\Msunyr$, and 0.2--0.7 in the most diffuse models, whose outer rays are density bounded; dense models can exceed case~B (\S\ref{sec:limits}).

Each emitting zone is given the velocity of the BH plus free fall (inside $R_{\rm acc}$) or that of the star plus the radial wind, distributed over twelve azimuths per ray and projected on the line of sight with the Gaia orbit (semi-amplitude of the giant $K_1=56.6$~\kms). The profiles cover $\pm16{,}000$~\kms, beyond the free-fall speed at $10^3\,R_g$ (13,400~\kms), on both the pre- and post-periastron branches, and are expressed as equivalent widths (EW) against the 5212~K photosphere.

\textsc{Cloudy} assumes ionization and thermal equilibrium at each point. This holds in the slowly moving wind but not in the inflow, which falls inward faster than it recombines, so that it carries its ionization state with it, and is heated by its own compression ($n\propto l^{-3/2}$) at a rate $(3/2)\,Pv_{\rm ff}/l$ that a static calculation does not contain. We estimate both effects with a simplified calculation along each ray, which follows the ionization of H and He and the thermal energy of each parcel as it falls at the free-fall speed $v$: \begin{equation} v\frac{dx_i}{dl}=\mathcal{I}_i-\mathcal{R}_i,\qquad \frac{3}{2}k\,v\frac{d(NT)}{dl}=\frac{\Gamma-\Lambda}{n_{\rm H}}+NkT\,v\frac{d\ln n_{\rm H}}{dl}, \end{equation} with $x_i$ the ion fractions, $\mathcal{I}_i$ and $\mathcal{R}_i$ the ionization and recombination rates, $\Gamma-\Lambda$ the radiative heating minus cooling, $N$ the number of particles per hydrogen nucleus, and the last term the compression work. The gas enters the inflow in equilibrium, and we recompute the attenuated spectrum and the flow in turn until the line luminosities converge to 0.3\%. The calculation uses standard rates \citep{verner96,shull85,voronov97,hui97,cen92,draine11} and is calibrated zone by zone so that, with the flow speed set to zero, it reproduces the \textsc{Cloudy} temperatures to 1\% and line luminosities to $10^{-3}$. We use the calculation only for the change that the flow produces: the \textsc{Cloudy} luminosity of each zone is multiplied by the ratio of the luminosities with and without the flow (capped at three, any excess being added as it is), and the line profiles are recomputed from the corrected zones. \S\ref{sec:robust} gives the result.

\section{Data and Methods}
\label{sec:data}
\subsection{Archival spectra and the X-ray observation}
The archival data (Table~\ref{tab:obs}) are two UVES spectra from November 2020 and May 2024 at $r/r_{\rm apo}=0.79$ and 1.00, and 21 ESPRESSO spectra from July--September 2024, of which we stack the 20 with S/N~$\ge20$. They show no line emission related to the BH. A Gaussian line at the BH velocity, fitted together with the stellar spectrum (UVES) or a local continuum (ESPRESSO stack), gives $3\sigma$ limits of EW(H$\alpha$) $<0.006$~\AA\ and EW(He\,II~4686) $<0.018$~\AA\ for a FWHM of 30~\kms, and 0.33 and 0.19~\AA\ for 400~\kms. The two instruments play different roles. We test H$\alpha$ only in the UVES spectra: the stellar H$\alpha$ line dominates the profile, and we can separate emission that moves with the BH from it only through the 25~\kms\ change in stellar velocity between the two UVES epochs, whereas the ESPRESSO epochs span 2~\kms. He\,II~4686 has no stellar counterpart, so we use the ESPRESSO stack, with its higher signal-to-noise ratio, with a local continuum fit. The Chandra observation (ObsID 29396) took place three days after the second UVES spectrum.

\begin{table*}[!tp]
\caption{Executed observations and simulated campaigns. $N$: number of epochs; $r/r_{\rm apo}$: separation in units of the apastron separation; $v_\star-\gamma$: radial velocity of the giant relative to the systemic velocity}
\label{tab:obs}
\centering
\begin{tabular}{lrp{7.6cm}cc}
\hline
 & $N$ & Dates & $r/r_{\rm apo}$ & $v_\star-\gamma$ [km\,s$^{-1}$]\\
\hline
Chandra (X-ray) & 1 & 2024 May 25 (12.6 ks) & 0.997 & \\
UVES & 2 & 2020 Nov 5; 2024 May 21 & 0.79; 0.997 & $-26$; $-1$\\
ESPRESSO & 21 & 2024 Jul--Sep & 0.99 & $0$ to $+2$\\
\textbf{Scenario A} & 14 & monthly, Apr--Oct 2026 and Apr--Oct 2027 & 0.89--0.68 & $+13$ to $+29$\\
\textbf{Scenario B} & 25 & monthly, Jun--Oct 2028; fortnightly, 2029 Apr 1--Nov 11; monthly, Apr--Jun 2030 & 0.55--0.156 & $-47$ to $+66$\\
\hline
\end{tabular}
\end{table*}

\subsection{Statistical test}
The predicted lines are neither narrow nor Gaussian. In fact, each line is the sum of two components (\S\ref{sec:lines}): a broad one from the inflow, which moves with the BH and falls freely toward it, with wings out to thousands of \kms, and a narrow one from the undisturbed wind, which moves with the giant and is tens of \kms\ wide. Their relative weight depends on the model and separation, so a Gaussian template at the BH velocity recovers a model-dependent fraction of the flux, about 1\% for inflow-dominated models. We therefore test each model with its own predicted profile. For each epoch, we take the profile predicted for that separation and side of the orbit (approaching or receding) and place its two components at the velocities implied by the measured stellar radial velocity for the giant and, through the orbit, for the BH. We then fit the profiles of all epochs to the spectra together as a single template scaled by one free amplitude $a$, which multiplies the predicted equivalent widths: $a=1$ means the spectra contain the line exactly as predicted, and $a=0$ means they contain no line. The fit includes the stellar spectrum and a quadratic baseline over $\pm900$~\kms\ (UVES, H$\alpha$), or the baseline alone (ESPRESSO stack, He\,II). The uncertainty $\sigma_a$ is the robust dispersion of the amplitudes the same fit returns when the template is placed at 300 line-free wavelengths, so it includes the data's correlated residuals.

The statistic measures how many sigmas the fitted amplitude falls short of the prediction. With $S_\ell=1/\sigma_{a,\ell}$ the significance at which a line of the predicted strength would have been detected, and $z_\ell=\hat a_\ell/\sigma_{a,\ell}$ that of the amplitude actually fitted,
\begin{equation}
\Delta\chi^2(\Mdw,\fEdd)=\sum_{\ell}\left[\max\left(S_\ell-\max(z_\ell,0),\,0\right)\right]^2 ,
\end{equation}
summed over H$\alpha$ and He\,II. The two truncations make this an upper-limit statistic: a negative fitted amplitude counts as zero, and a model fainter than the fitted amplitude is not penalized. They matter for H$\alpha$, where templates much broader than the fitting window, or that follow the star, pick up residuals of the stellar line, which varies between the two UVES epochs and is absent from the line-free comparison regions; such templates return $|z_\ell|$ up to 5 for models far below the noise ($S_\ell<0.01$) and would otherwise exclude or favor models at random. No model is detected: for the four baseline models that the data could have seen ($S>1$) the fitted amplitudes are within $1.4\sigma$ of zero. We interpolate $\log S$ and $z$ between grid nodes and use $\Delta\chi^2=6.18$ for the 95\% region in two parameters, and 2.71 (95\%) and 9 ($3\sigma$) for one-sided limits on $\Mdw$ at fixed $\fEdd$.

\subsection{Campaign scenarios}\label{sec:scenarios}
We simulate two campaigns that differ only in when the spectra are taken (Table~\ref{tab:obs}, Fig.~\ref{fig:cartoon}a). Both assume the quality of the 2024 UVES spectra: a signal-to-noise ratio of 184 per 1~\kms\ pixel at each epoch, the systematics measured in those spectra, and the observed epoch-to-epoch changes of the stellar line cores. This corresponds to one 45-min exposure per epoch on an 8-m telescope at a resolving power of 87,000 around H$\alpha$ and 65,000 around He\,II~4686. The star is observable from April to October.

\paragraph{Scenario~A: monitoring away from periastron} One epoch per month from April to October in 2026 and 2027, 14 epochs at $r/r_{\rm apo}=0.89$--0.68. The BH is then on the same part of the orbit as in the archival data and the Chandra observation, and the star's velocity changes by only 16~\kms. It serves as a control for scenario~B.

\paragraph{Scenario~B: periastron campaign} Twenty-five epochs: monthly from June to October 2028, fortnightly from April 1, 2029 to November 11, 2029 (17 epochs), and monthly from April to June 2030. The separation falls to $r/r_{\rm apo}=0.156$, and the star's velocity sweeps over 113~\kms. From Chile, the star has less than 2~h of dark time per night from mid-October 2029, so the epochs closest to periastron need a northern 8--10-m telescope with a spectrograph that covers both lines, such as Keck/HIRES.

For each model, we inject the predicted profiles, interpolated in $\log r$ and placed at the BH and star velocities from the ephemeris refined with the ground-based radial velocities (periastron on November 13, 2029), into simulated spectra and recover them with the estimator used in the archival test. The detection significance, $\sigma_{\rm A}$ or $\sigma_{\rm B}$, is the inverse of the robust dispersion of the recovered amplitude over 1500 noise realizations.

\section{Results}\label{sec:results}
This section presents the predicted line strengths and profiles across the model grid (\S\ref{sec:lines}), the constraints that the archival UVES and ESPRESSO spectra place on the models (\S\ref{sec:constraints}), the detectability of the lines in the two simulated campaigns (\S\ref{sec:campaigns}), and the sensitivity of these results to the model assumptions (\S\ref{sec:robust}).

\subsection{Line strengths and profiles}\label{sec:lines}

\begin{figure*}[!tp]
\centering\includegraphics[width=\textwidth]{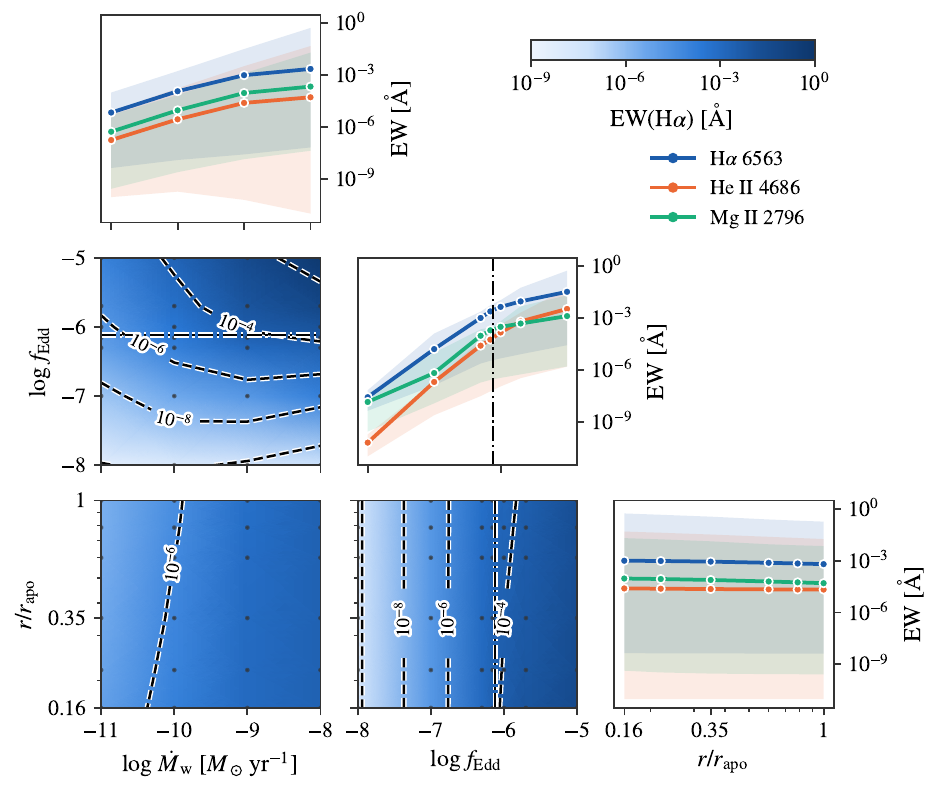}
\caption{Predicted equivalent widths over the three model parameters (wind mass-loss rate $\Mdw$, Eddington ratio $\fEdd$, relative separation $r/r_{\rm apo}$) for the slow wind ($v_\infty=30$~\kms, $r_{\rm hot}=10^3\,R_g$). Off-diagonal panels: EW(H$\alpha$) (color) and EW(He\,II~4686) (dashed contours, labeled in \AA) on two-parameter slices through the reference model ($\Mdw=10^{-9}\,\Msunyr$, $\fEdd=5\times10^{-7}$, periastron); dots mark the grid models. Diagonal panels: EW of H$\alpha$, He\,II~4686 and Mg\,II~2796 along one parameter through the reference model (lines) and their range over the other two parameters (bands). Dash-dotted vertical line: the bound on $\fEdd$ from the Chandra observation of \citet{cappelluti24} (\S\ref{sec:model}).}
\label{fig:ew}
\end{figure*}

The Eddington ratio primarily sets the line strengths, with the wind mass-loss rate playing a secondary role, as shown in Fig.~\ref{fig:ew}. By contrast, the separation has little effect on the line strengths. 

At $\Mdw=10^{-8}\,\Msunyr$ the periastron EW(H$\alpha$) rises from $6.6\times10^{-8}$~\AA\ at $\fEdd=10^{-8}$ to $5.3\times10^{-3}$~\AA\ at $7\times10^{-7}$, just below the Chandra bound, and 0.54~\AA\ at $10^{-5}$. The mass-loss rate matters less, and how much depends on $\fEdd$. At $\fEdd=10^{-5}$ the ionized region extends into the wind and EW(H$\alpha$) grows from $10^{-4}$ to 0.54~\AA\ over the three decades of $\Mdw$; at $10^{-8}$ the source is starved of photons and the same range changes it only from $4.5\times10^{-9}$ to $6.6\times10^{-8}$~\AA.
The orbit modulates the lines only modestly. When the ionized region reaches the wind, the lines brighten toward periastron; over the 28 $(\Mdw,\fEdd)$ combinations, the periastron-to-apastron ratio of EW(H$\alpha$) is at most 4.1. In the densest models (e.g.\ $\Mdw\ge10^{-9}\,\Msunyr$ with $\fEdd\le10^{-7}$, or $\Mdw=10^{-8}\,\Msunyr$ with $\fEdd\le10^{-6}$) all ionizing photons are absorbed in the free-fall cusp around the BH, and since the capture rate does not depend on phase, the EW changes by less than 5\% between apastron and periastron.

The emission has two components with very different widths, and which one dominates determines whether a line can be detected. Gas inside the accretion radius moves with the BH as it falls toward it, so that its emission forms a cusp at the BH velocity, with wings extending to thousands of \kms. Gas in the undisturbed wind moves with the giant at $<v_\infty$ and emits a core only tens of \kms\ wide, offset from the BH velocity (because of the orbital motion). The balance between the two depends on where the ionizing photons are absorbed. In nine of the 28 $(\Mdw,\fEdd)$ combinations, they are all absorbed in the cusp before they reach the wind, and the cusp essentially emits all the H$\alpha$; we call these models cusp-dominated. In the other nineteen, the wind-dominated models, the wind contributes between a few percent and about half of the line. 

This matters because the archival limits were derived for narrow Gaussian lines at the BH velocity. Such a template recovers roughly half of the H$\alpha$ flux of a wind-dominated model (17--65\%), but only about one percent of that of a cusp-dominated one, whose flux sits in the broad wings and is absorbed into the continuum of the fit (Fig.~\ref{fig:onarchival}, bottom). The model with $\Mdw=10^{-8}\,\Msunyr$ and $\fEdd=7\times10^{-7}$ makes the point: its total equivalent width amounts to 0.84 of the tabulated narrow-line limit, yet its own profile would have registered at $0.06\sigma$. 

H$\alpha$ is the most constraining line. He\,II~4686 never exceeds a tenth of the H$\alpha$ equivalent width, falls to a few percent near the Chandra bound, and nearly vanishes in the densest models. In addition, because the archival limit on He\,II is also weaker, H$\alpha$ constrains the grid everywhere. The ultraviolet does not change this. Below the Chandra bound, Mg\,II~2796 reaches at most $5\times10^{-18}$~erg\,s$^{-1}$\,cm$^{-2}$ at Earth, two orders of magnitude below what HST/STIS detects in a few orbits and below the chromospheric Mg\,II of the giant itself. He\,II~1640 is brighter, but it forms in the inflow, where advection suppresses it (\S\ref{sec:robust}), and is spread over thousands of \kms. Interstellar absorption removes Ly$\alpha$ within $\pm1500$~\kms\ of the line center.

\subsection{Constraints from the archival spectra}\label{sec:constraints}

\begin{figure*}[!tp]
\centering\includegraphics[width=\textwidth]{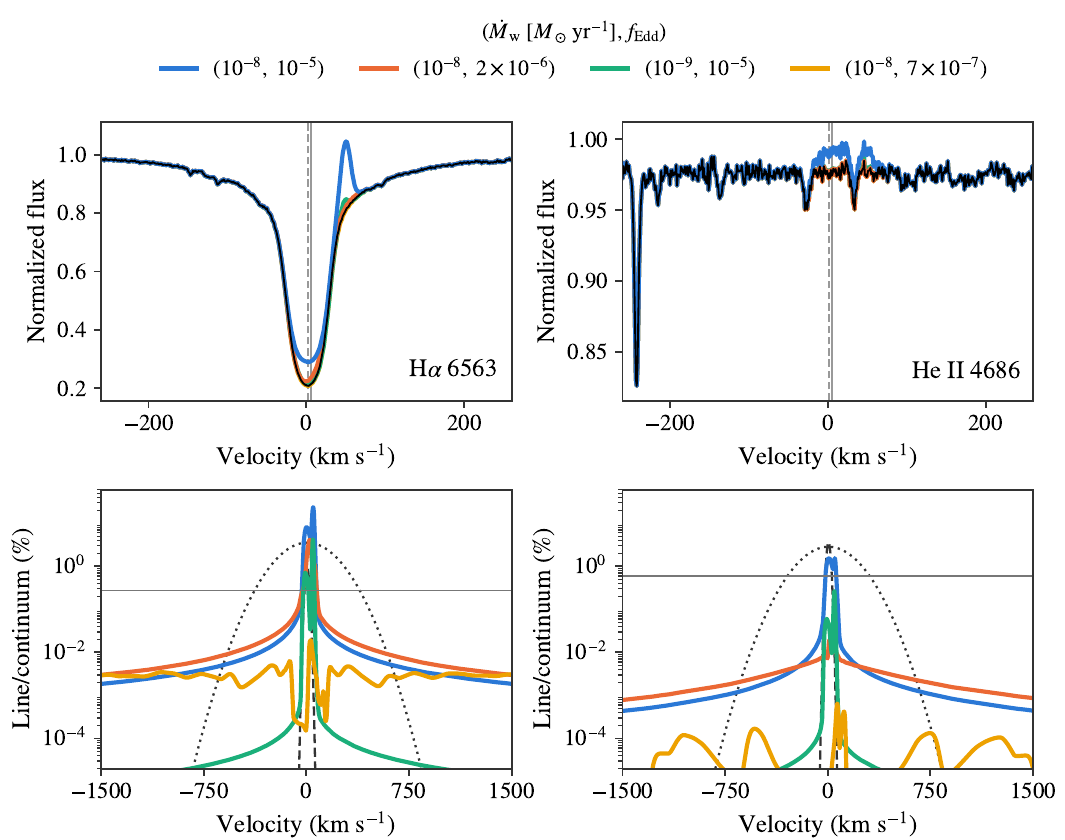}
\caption{Model line profiles on the archival spectra. Top: the stack of the 20 ESPRESSO spectra of July--September 2024 ($r/r_{\rm apo}=0.99$; black) around H$\alpha$ (left) and He\,II~4686 (right), in the systemic frame, and the same spectrum with the apastron profile of four baseline models added (colors; $\Mdw$ in $\Msunyr$; a model is visible only where it departs from the data). Vertical lines: velocity of the BH (solid) and of the star (dashed). Bottom: the model profiles alone, in percent of the continuum, with their far wings smoothed for display. Black: Gaussian lines at the BH velocity that carry the archival $3\sigma$ limits (H$\alpha$: UVES; He\,II: ESPRESSO) for FWHM 30~\kms\ (dashed) and 400~\kms\ (dotted); horizontal line: $1\sigma$ noise per 1~\kms\ pixel of the stack. The blue, orange and green models are excluded by the archival test; the yellow one, the brightest model below the Chandra bound, is allowed. A matched filter with the exact profile would recover the four models from the stack at 340, 68, 45 and $0.3\sigma$ in H$\alpha$ if the stellar spectrum were known.}
\label{fig:onarchival}
\end{figure*}

Of the 28 baseline $(\Mdw,\fEdd)$ combinations, the archival spectra exclude three at 95\% confidence: $\Mdw=10^{-8}\,\Msunyr$ with $\fEdd=2\times10^{-6}$, which the data would have shown at $4.0\sigma$, and with $\fEdd=10^{-5}$ ($66\sigma$), and $\Mdw=10^{-9}\,\Msunyr$ with $\fEdd=10^{-5}$ ($11\sigma$). The next one, $\Mdw=10^{-9}\,\Msunyr$ with $\fEdd=2\times10^{-6}$, would have been seen at $1.1\sigma$, and all others have $\Delta\chi^2\le0.2$. H$\alpha$ drives the exclusion; in He\,II the brightest model reaches $S=1.7\sigma$. Figure~\ref{fig:onarchival} shows why we exclude the three models and keep the others. The excluded models add to the ESPRESSO stack H$\alpha$ components that reach 24\%, 4.1\% and 4.2\% of the continuum, against a noise of 0.28\% per pixel, whereas the brightest model below the Chandra bound ($\Mdw=10^{-8}\,\Msunyr$, $\fEdd=7\times10^{-7}$) stays below 0.02\% at every velocity, its EW being spread over the free-fall cusp. A test with Gaussian templates at the BH velocity excludes the two models with $\Mdw=10^{-8}\,\Msunyr$ but misses $\Mdw=10^{-9}\,\Msunyr$ with $\fEdd=10^{-5}$, whose narrow component lies outside a 30~\kms\ template centered on the BH.

The resulting limits are $\Mdw<2\times10^{-9}\,\Msunyr$ (95\%; $6\times10^{-9}$ at $3\sigma$) for $\fEdd=2\times10^{-6}$ and $\Mdw<2.6\times10^{-10}\,\Msunyr$ (95\%; $4\times10^{-10}$ at $3\sigma$) for $\fEdd=10^{-5}$. For $\fEdd\le10^{-6}$ the grid gives no limit. Since the lines depend much more steeply on $\fEdd$ than on $\Mdw$, the boundary of the excluded region is nearly horizontal (Fig.~\ref{fig:plane}a), and it lies entirely above the Chandra bound: at $\fEdd=10^{-6}$ no model would have been seen at more than $0.4\sigma$.

These limits hold for the baseline configuration. Enlarging the non-emitting hot flow to $r_{\rm hot}=10^5\,R_g$ removes only the innermost part of the cusp and leaves them unchanged. A fast wind, by contrast, removes them: its lines are fainter (\S\ref{sec:robust}), no fast-wind model is excluded, and even the brightest, $\Mdw=10^{-8}\,\Msunyr$ with $\fEdd=10^{-5}$, would have been seen at only $1.9\sigma$. For that model, the fit actually returns a positive amplitude, $1.5\pm0.5$ times the prediction. We do not read this as a detection. In the fast-wind models, the emission comes from the wind, so the template moves with the star and responds to any change in the stellar H$\alpha$ line between the two UVES epochs; templates of fast-wind models predicted far below the noise return amplitudes up to $+5.3\sigma$, which can only be stellar variability. The archival limits therefore apply to a slow wind. Testing the fast-wind models requires more epochs or an independent handle on the stellar line's variability.

\subsection{Predictions for the two campaigns}\label{sec:campaigns}

\begin{figure*}[!tp]
\centering\includegraphics[width=\textwidth]{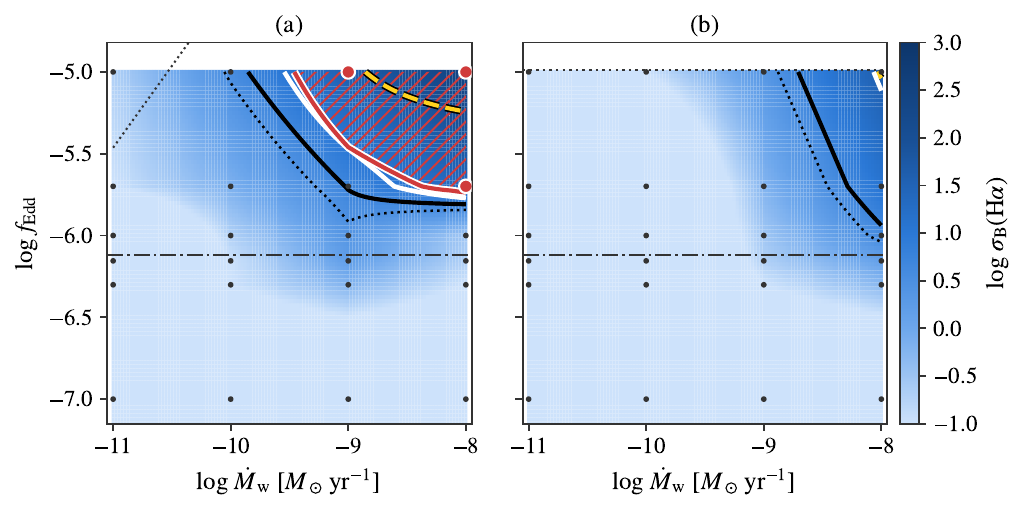}
\caption{Archival constraints and campaign detectability for the slow (a) and fast (b) wind, both with $r_{\rm hot}=10^3\,R_g$. Color: detection significance of H$\alpha$ in scenario~B, $\sigma_{\rm B}$, interpolated between the grid models (dots); black contours: $\sigma_{\rm B}=3$ (dotted) and 5 (solid); yellow dashed: $\sigma_{\rm B}=5$ for He\,II~4686; white: $\sigma_{\rm A}=5$ for H$\alpha$ in scenario~A. Red hatched: region excluded by the archival spectra at 95\%, with the excluded grid models circled; no fast-wind model is excluded. Dash-dotted: Chandra bound on $\fEdd$ at apastron (2024 May). Dotted in (b): the Chandra bound carried to periastron for a constant $\epsilon$, using the 13.6-fold rise of the capture rate; for the slow wind the capture rate does not change with phase. Gray dotted line in (a): $\epsilon=1$ ($\fEdd=\dot m_{\rm BHL}$), the Eddington ratio if all the captured gas reached the black hole; the adopted spectrum, in which the flow loses gas to an outflow inside $r_{\rm hot}$ (\S\ref{sec:model}), cannot reach it. The grid extends to $\fEdd=10^{-8}$, where no line is detectable; Table~\ref{tab:pred} provides the values at the nodes.}
\label{fig:plane}
\end{figure*}

\begin{table*}[!tp]
\caption{Predictions for the slow-wind grid ($v_\infty=30$~\kms, $r_{\rm hot}=10^3\,R_g$): EWs at periastron, detection significance
in scenario~A ($\sigma_{\rm A}$, H$\alpha$) and scenario~B ($\sigma_{\rm B}$, H$\alpha$ and He\,II) from 1500 realizations, and status in
the archival test}\label{tab:pred}
\centering\begin{tabular}{llcccccc}
\hline
$\dot M_{\rm w}$ & $f_{\rm Edd}$ & EW(H$\alpha$) & EW(He\,II) & $\sigma_{\rm A}$(H$\alpha$) & $\sigma_{\rm B}$(H$\alpha$) & $\sigma_{\rm B}$(He\,II) & archival\\
{[M$_\odot$\,yr$^{-1}$]} & & [\AA] & [\AA] & & & & 95\%\\
\hline
$10^{-11}$ & $10^{-8}$ & $4.5\times10^{-9}$ & $8.8\times10^{-11}$ & 0.0 & 0.0 & 0.0 & allowed\\
$10^{-11}$ & $10^{-7}$ & $5.1\times10^{-7}$ & $3.4\times10^{-9}$ & 0.0 & 0.0 & 0.0 & allowed\\
$10^{-11}$ & $5\times10^{-7}$ & $6.8\times10^{-6}$ & $1.8\times10^{-7}$ & 0.0 & 0.0 & 0.0 & allowed\\
$10^{-11}$ & $7\times10^{-7}$ & $1.0\times10^{-5}$ & $3.3\times10^{-7}$ & 0.0 & 0.0 & 0.0 & allowed\\
$10^{-11}$ & $10^{-6}$ & $1.4\times10^{-5}$ & $5.9\times10^{-7}$ & 0.0 & 0.0 & 0.0 & allowed\\
$10^{-11}$ & $2\times10^{-6}$ & $2.8\times10^{-5}$ & $1.5\times10^{-6}$ & 0.0 & 0.1 & 0.0 & allowed\\
$10^{-11}$ & $10^{-5}$ & $9.8\times10^{-5}$ & $7.9\times10^{-6}$ & 0.1 & 0.2 & 0.0 & allowed\\
$10^{-10}$ & $10^{-8}$ & $1.2\times10^{-8}$ & $1.8\times10^{-10}$ & 0.0 & 0.0 & 0.0 & allowed\\
$10^{-10}$ & $10^{-7}$ & $3.6\times10^{-6}$ & $1.0\times10^{-7}$ & 0.0 & 0.0 & 0.0 & allowed\\
$10^{-10}$ & $5\times10^{-7}$ & $1.2\times10^{-4}$ & $2.8\times10^{-6}$ & 0.0 & 0.1 & 0.0 & allowed\\
$10^{-10}$ & $7\times10^{-7}$ & $1.9\times10^{-4}$ & $6.5\times10^{-6}$ & 0.1 & 0.1 & 0.0 & allowed\\
$10^{-10}$ & $10^{-6}$ & $2.6\times10^{-4}$ & $1.3\times10^{-5}$ & 0.2 & 0.2 & 0.0 & allowed\\
$10^{-10}$ & $2\times10^{-6}$ & $5.9\times10^{-4}$ & $3.7\times10^{-5}$ & 0.4 & 0.7 & 0.0 & allowed\\
$10^{-10}$ & $10^{-5}$ & $1.7\times10^{-3}$ & $1.7\times10^{-4}$ & 1.3 & 3.5 & 0.4 & allowed\\
$10^{-9}$ & $10^{-8}$ & $2.6\times10^{-8}$ & $6.1\times10^{-11}$ & 0.0 & 0.0 & 0.0 & allowed\\
$10^{-9}$ & $10^{-7}$ & $1.6\times10^{-5}$ & $2.0\times10^{-7}$ & 0.0 & 0.0 & 0.0 & allowed\\
$10^{-9}$ & $5\times10^{-7}$ & $9.9\times10^{-4}$ & $2.5\times10^{-5}$ & 0.1 & 0.6 & 0.0 & allowed\\
$10^{-9}$ & $7\times10^{-7}$ & $2.4\times10^{-3}$ & $5.5\times10^{-5}$ & 0.2 & 1.5 & 0.0 & allowed\\
$10^{-9}$ & $10^{-6}$ & $4.3\times10^{-3}$ & $1.4\times10^{-4}$ & 0.5 & 2.4 & 0.1 & allowed\\
$10^{-9}$ & $2\times10^{-6}$ & $9.0\times10^{-3}$ & $6.3\times10^{-4}$ & 2.3 & 5.3 & 0.4 & allowed\\
$10^{-9}$ & $10^{-5}$ & $3.2\times10^{-2}$ & $3.3\times10^{-3}$ & 25.0 & 39.5 & 3.5 & excluded\\
$10^{-8}$ & $10^{-8}$ & $6.6\times10^{-8}$ & $9.8\times10^{-12}$ & 0.0 & 0.0 & 0.0 & allowed\\
$10^{-8}$ & $10^{-7}$ & $1.2\times10^{-4}$ & $3.8\times10^{-8}$ & 0.0 & 0.0 & 0.0 & allowed\\
$10^{-8}$ & $5\times10^{-7}$ & $2.3\times10^{-3}$ & $5.2\times10^{-5}$ & 0.0 & 0.1 & 0.0 & allowed\\
$10^{-8}$ & $7\times10^{-7}$ & $5.3\times10^{-3}$ & $1.5\times10^{-4}$ & 0.1 & 0.2 & 0.0 & allowed\\
$10^{-8}$ & $10^{-6}$ & $1.1\times10^{-2}$ & $4.0\times10^{-4}$ & 0.1 & 0.3 & 0.0 & allowed\\
$10^{-8}$ & $2\times10^{-6}$ & $5.5\times10^{-2}$ & $2.0\times10^{-3}$ & 17.1 & 25.2 & 0.1 & excluded\\
$10^{-8}$ & $10^{-5}$ & $5.4\times10^{-1}$ & $4.8\times10^{-2}$ & 107.5 & 301.6 & 40.2 & excluded\\
\hline
\end{tabular}
\end{table*}

Figure~\ref{fig:plane} and Table~\ref{tab:pred} show how significantly the two campaigns would recover the predicted lines. Scenario~A detects H$\alpha$ at $\ge5\sigma$ only in the three models that the archival spectra already exclude ($\sigma_{\rm A}=108$, 25 and 17). It is 1.4 to 4.3 times as sensitive as the archival data, but its $5\sigma$ contour runs along the archival 95\% boundary: the best allowed model, $\Mdw=10^{-9}\,\Msunyr$ with $\fEdd=2\times10^{-6}$, reaches $2.3\sigma$, and no model with $\fEdd\le10^{-6}$ exceeds $0.5\sigma$. A non-detection would lower the limits on $\Mdw$ by a factor of two to three, to $6.5\times10^{-10}$ and $1.2\times10^{-10}\,\Msunyr$ at $\fEdd=2\times10^{-6}$ and $10^{-5}$. The gain is small because the separation is as large as in the archival spectra and the stellar velocity hardly changes, making it hard to separate a line that moves with the BH from the star. For the fast wind, scenario~A detects only the brightest model ($6.1\sigma$).

Scenario~B observes the BH at separations down to a sixth of those of scenario~A while the star sweeps over 113~\kms. For the slow wind it detects the allowed model $\Mdw=10^{-9}\,\Msunyr$ with $\fEdd=2\times10^{-6}$ at $5.3\sigma$; $\Mdw=10^{-10}\,\Msunyr$ with $\fEdd=10^{-5}$ gives $3.5\sigma$, and $\Mdw=10^{-9}\,\Msunyr$ with $\fEdd=10^{-6}$, just above the Chandra bound, $2.4\sigma$. Below the bound, none of the 16 models exceeds $1.5\sigma$. The model with $\Mdw=10^{-8}\,\Msunyr$ and $\fEdd=7\times10^{-7}$ has a periastron EW(H$\alpha$) of $5.3\times10^{-3}$~\AA\ but gives only $0.2\sigma$: it is cusp-dominated, so its flux is spread over thousands of \kms\ and hardly changes with phase. A non-detection would give $\Mdw\lesssim2.6\times10^{-10}\,\Msunyr$ at $\fEdd=2\times10^{-6}$ and $\lesssim5.5\times10^{-11}\,\Msunyr$ at $10^{-5}$ (95\%), eight and five times below the archival limits. He\,II reaches $5\sigma$ only in the brightest excluded model; no allowed model exceeds $0.4\sigma$, so H$\alpha$ without He\,II is the expected signature.

For the fast wind, the rising capture rate makes scenario~B sensitive to a region that neither the archival spectra nor scenario~A reach (Fig.~\ref{fig:plane}b). It detects H$\alpha$ at 11.8 and $40.6\sigma$ for $\Mdw=10^{-8}\,\Msunyr$ with $\fEdd=2\times10^{-6}$ and $10^{-5}$, and at $4\sigma$ with $\fEdd=10^{-6}$; its $5\sigma$ contour reaches $\fEdd\simeq1.2\times10^{-6}$ at $\Mdw=10^{-8}\,\Msunyr$ and $\Mdw\simeq2\times10^{-9}\, \Msunyr$ at $\fEdd=10^{-5}$. The fast-wind models keep $\fEdd$ fixed along the orbit. If $\epsilon$ rather than $\fEdd$ were constant, $\fEdd$ at the 17 epochs of 2029 ($r/r_{\rm apo}=0.32$--0.16) would be 0.4--1 times its periastron value, so the fixed-$\fEdd$ models are adequate for them and overestimate the signal in 2028 and 2030.

\begin{figure*}[!tp]
\centering\includegraphics[width=0.86\textwidth]{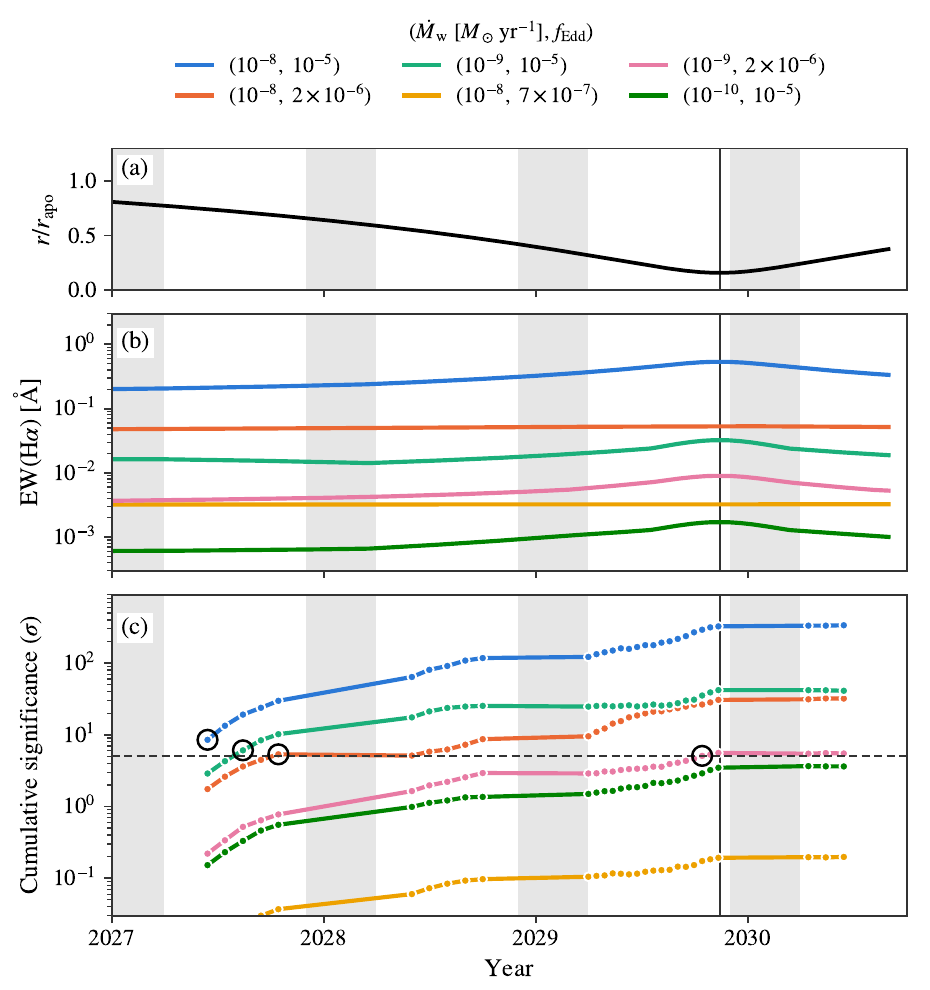}
\caption{Time evolution of the predicted H$\alpha$ signal and of its detectability for six slow-wind models (colors). (a) Orbital separation for the refined ephemeris; the vertical line marks periastron (November 13, 2029), and the gray bands mark the months when the star is not observable. (b) Predicted EW(H$\alpha$). (c) Cumulative detection significance after each epoch (dots) of a campaign made of the 2027 season of scenario~A followed by the 25 epochs of scenario~B. The dashed line is $5\sigma$, and the circles mark the first epoch at which a model exceeds it. The archival data exclude the blue, orange, and green models; the yellow one is the brightest model below the Chandra bound. The final significance levels, in legend order (by columns), are 337, 32, 41, 0.2, 5.5, and $3.6\sigma$.}
\label{fig:snrt}
\end{figure*}

Figure~\ref{fig:snrt} follows the signal through the 2027 season of scenario~A and the epochs of scenario~B. Between apastron and periastron, the EW of the six models shown grows by at most a factor of 2.8, and stays flat for the cusp-dominated ones, so significance builds with the number of epochs and the star's velocity excursion, not at periastron itself. The model with $\Mdw=10^{-9}\,\Msunyr$ and $\fEdd=2\times10^{-6}$ crosses $5\sigma$ on October 14, 2029, a month before periastron, and the three excluded models would already cross it in 2027.

\subsection{Sensitivity to the model assumptions}
\label{sec:robust}
Advection and compression (\S\ref{sec:model}) change the inflow state considerably but have little effect on the lines. The inflow carries $\dot N=7.1\times10^{37}\,(\Mdw/10^{-11}\,\Msunyr)$ hydrogen nuclei per second toward the BH. Where $Q({\rm H})<\dot N$, i.e.\ below $\fEdd=1.4\times10^{-7}$, $2.7\times10^{-7}$, $5.6\times10^{-7}$ and $1.3\times10^{-6}$ for $\Mdw=10^{-11}$, $10^{-10}$, $10^{-9}$ and $10^{-8}\,\Msunyr$, the source cannot ionize the gas that advection supplies, and the inflow stays neutral down to $r_{\rm hot}$. Its He\,II emission disappears, and it radiates the compression work ($\simeq17$~eV per hydrogen nucleus) in collisionally excited hydrogen lines, about 1\% of it in H$\alpha$. This H$\alpha$ floor is spread over 400--5000~\kms\ and is undetectable. Where the inflow is ionized, it heats up to $\sim10^8$~K near $r_{\rm hot}$ and its own lines are suppressed. The three excluded models keep 92--100\% of their equilibrium H$\alpha$ and remain excluded, the archival limits change by less than 10\%, and the campaign significances of the allowed models change by less than 2\%. The one exception is $\Mdw=10^{-8}\,\Msunyr$ with $\fEdd=10^{-6}$, where $Q({\rm H})\simeq\dot N$ and the iteration alternates between an ionized and a neutral inflow, so that its corrected EW is uncertain by tens of percent. Below the Chandra bound the brightest corrected periastron EW(H$\alpha$) is $3.4\times10^{-3}$~\AA, against $5.3\times10^{-3}$~\AA\ in equilibrium.

The fast wind has a lower density at a given $\Mdw$ and the BH captures a smaller fraction of it, so its H$\alpha$ EWs are a median 0.23 of the baseline values (16th--84th percentiles 0.20--0.50). A hot flow of $10^5\,R_g$ removes the innermost part of the cusp; it leaves most models unchanged (median 0.99) but reduces the most cusp-dominated one by more than a factor of twenty. The baseline is therefore the most favorable of the four configurations.

\section{Discussion}\label{sec:discussion}
\subsection{What the lines measure}\label{sec:xray}

The X-ray, radio, and optical bands probe different aspects of the problem. The 2--10~keV luminosity comes from the hot flow itself, and the radio luminosity comes from an outflow or jet; both trace the flow's state, and both limits come from single epochs near apastron. The recombination lines count the ionizing photons the surrounding gas absorbs, and their strength and profile depend on the gas density and velocity, that is, on $\Mdw$ and how the wind is captured. They can also be observed at any phase.

Where the bands overlap, they agree. Within the ADAF model, the region excluded by the archival spectra lies at Eddington ratios that the Chandra observation also excludes (Fig.~\ref{fig:plane}a). The two limits sample different parts of the spectrum: for the ADAF the Chandra limit corresponds to $Q({\rm H})<1.8\times10^{40}$~s$^{-1}$, while the archival spectra exclude $Q({\rm H})\ge2.4\times10^{41}$~s$^{-1}$ if $\Mdw\ge2\times10^{-9}\,\Msunyr$. For a spectrum with more than 14 times as many ionizing photons per unit 2--10~keV luminosity as the ADAF, the lines would give the tighter limit.

The radio limit constrains how observable the lines are. If Gaia~BH3 followed the fundamental plane of black-hole activity, the VLA limit would imply $L_X\lesssim10^{27}$~erg\,s$^{-1}$ (\S1), i.e., $\fEdd\approx5\times10^{-8}$ at apastron (Table~\ref{tab:sed}). No line would then be detectable in either campaign, even for a fast wind, whose periastron $\fEdd$ would be at most $7\times10^{-7}$ (scenario~B reaches $1.3\sigma$ there). The radio luminosity scatters about the fundamental plane by 0.88~dex \citep{merloni03}, so the VLA limit lies $1.6\sigma$ below the relation. A line detected near periastron would therefore also show that Gaia~BH3 is radio-quiet relative to the fundamental plane, or that its accretion state changes along the orbit.

Monitoring over the orbit observed so far (scenario~A) is 1.4--4.3 times as sensitive as the archival spectra but reaches $5\sigma$ only for models they already exclude. Its main use is to characterize the star: it would measure epoch-to-epoch changes in the stellar H$\alpha$ profile that limit tests of a fast wind (\S\ref{sec:constraints}) and provide stellar templates for a later campaign.

Near periastron, the emissivity's response to the orbit depends on the wind. For a slow wind, the BH captures the same fraction of the wind at every phase; if $\epsilon$ is also constant, $\fEdd$ at periastron equals that of May 2024, and scenario~B detects nothing. A detection would then show that the accretion state changes along the orbit, with $\fEdd\gtrsim2\times10^{-6}$ at periastron and $\Mdw\ge10^{-9}\,\Msunyr$. For a fast wind, the capture rate rises 13.6-fold toward periastron, and the apastron limit allows $\fEdd$ up to $10^{-5}$ there (Fig.~\ref{fig:plane}b); scenario~B then measures the response of the flow and of the gas around it to a known change in the supply.

A detection would measure what a luminosity in another band cannot. The line flux, combined with $\fEdd$ from an X-ray observation at the same epoch, gives $\Mdw$; the profile separates the gas falling onto the BH from the wind; the velocity curve shows whether the line follows the BH or the star; and He\,II/H$\alpha$ measures the hardness of the ionizing spectrum. Together they give $\epsilon$.

\subsection{How much of the wind reaches the BH}
\label{sec:eff}
For the slow wind the line $\epsilon=1$ in Fig.~\ref{fig:plane}a ($\fEdd=\dot m_{\rm BHL}$) lies above the Chandra limit for every $\Mdw$ of the grid, which gives the X-ray limit: $\epsilon\lesssim0.2$ at $\Mdw=10^{-11}\,\Msunyr$ and $\lesssim2\times10^{-4}$ at $10^{-8}\,\Msunyr$. The lines constrain $\epsilon$ independently of the X-ray data. The three models excluded by the archival spectra have $\epsilon=6\times10^{-4}$, $3\times10^{-3}$ and 0.03, so if the giant loses $\gtrsim2\times10^{-9}\,\Msunyr$, less than $\sim3\times10^{-3}$ of the captured gas reaches the BH. Because the hot flow loses most of its gas to an outflow on the way in (\S\ref{sec:model}), it can take in up to 22 times more: for $\Mdw\lesssim5\times10^{-11}\,\Msunyr$, close to the Reimers rate, the Chandra limit allows the whole captured wind to enter it at $r_{\rm hot}$. Neither band probes accretion from a Reimers wind, and below the Chandra limit a slow wind produces no line detectable in either campaign or in the ultraviolet (\S\ref{sec:lines}). In that regime the wind has to be measured independently of the accretion, in absorption against the giant's ultraviolet spectrum. An estimate for our velocity law, with Mg\,II and Fe\,II as the dominant ions, gives optical depths $\tau\simeq2\,(\Mdw/10^{-11}\,\Msunyr)$ in Mg\,II~k and $\tau\simeq0.3\,(\Mdw/10^{-10}\,\Msunyr)$ in Fe\,II~2374 ($\tau\propto \Mdw/v_\infty^2$), so together these lines remain unsaturated for $\Mdw\sim10^{-12}$--$10^{-9}\,\Msunyr$. Combined with an X-ray $\fEdd$, such a measurement would give $\epsilon$ directly, the quantity on which the predicted emission of the Galaxy's dormant black holes depends.

\subsection{Limitations}\label{sec:limits}
All predictions depend on the ADAF spectrum, which sets both the line strengths and the Eddington ratio that corresponds to the Chandra limit. Within the model the advected fraction matters at the level of tens of percent: with $f=1$ instead of the self-consistent values, $Q({\rm H})$ would be 13--45\% lower and the H$\alpha$ EWs 4--35\% lower, while the Chandra bound would move only to $7.7\times10^{-7}$. A different ionizing-photon-to-X-ray ratio would shift the optical and X-ray bounds relative to each other (\S\ref{sec:xray}).

The wind is assumed to be smooth and spherical. Clumping would raise the wind-dominated lines by the clumping factor $\langle n^2\rangle/\langle n\rangle^2$, and a focused wake would add narrow, phase-dependent emission that our estimator recovers efficiently; in these respects, the predicted lines are lower limits. The rays do not exchange diffuse radiation; we exclude the giant's own radiation field, which can photoionize excited hydrogen close to the star, and neglect occultation by the star. For a fast wind, adding wind and orbital speeds in quadrature overestimates the capture before periastron and underestimates it afterward.

Equilibrium is a good approximation for the wind. In ionization-bounded models, every ionizing photon absorbed in the wind is followed by a recombination in the same gas within a recombination time, so advection only displaces the emission downstream; in density-bounded models, the wind is fully ionized, and its recombination emission follows the local density. The response to the changing separation is delayed by at most $t_{\rm rec}\simeq4\times10^{12}\,{\rm s}/n_e$, i.e., weeks at the base of the wind for $\Mdw\ge10^{-9}\,\Msunyr$.

In the dense, photon-starved models, \textsc{Cloudy} returns more H$\alpha$ than case~B allows for the absorbed ionizing photons, up to 24 times more for $\Mdw=10^{-8}\,\Msunyr$ and $\fEdd=10^{-7}$. The energy budget is respected ($L({\rm H\alpha})$ is at most 7\% of the ionizing luminosity), and the heating and cooling budgets of the axis rays show where the excess comes from. Collisionally excited hydrogen lines carry 63--94\% of the cooling in all models, but in the models above case~B a large share of the heating comes from free--free absorption of the low-frequency ADAF continuum (57\% in the model with the largest excess), which a photon-counting estimate cannot include. This excess depends on the radio-to-infrared part of the spectrum and occurs in the models where the equilibrium assumption is least secure; none of them is detectable in the archival data or in either campaign (Table~\ref{tab:pred}).

\section{Conclusions} 

Gaia~BH3, a $32.7\,M_\odot$ black hole on an 11.6-yr orbit around a metal-poor giant 590~pc away, is a rare case in which we can compute an accreting black hole's fuel supply directly. In fact, the orbit is known, and the supply is the wind of a star whose spectrum can be studied. Two quantities remain unknown: (i) the mass-loss rate of the giant, $\Mdw$, and (ii) the fraction $\epsilon$ of the captured gas that reaches the black hole. The Chandra non-detection in May 2024 constrains their product at one epoch, apastron, requiring $\fEdd<7.6\times10^{-7}$ for an ADAF spectrum, but it does not constrain the wind. 

We investigate whether optical lines from gas photoionized by the hot flow can separate the two, particularly with observations near the periastron of November 2029. We computed the H$\alpha$ and He\,II~4686 emission of the wind and of the inflow along the orbit with a grid of 168 \textsc{Cloudy} models spanning $\Mdw=10^{-11}$--$10^{-8}\,\Msunyr$ and $\fEdd=10^{-8}$--$10^{-5}$, with variants for a fast wind. We derived line profiles from the gas kinematics, tested each model against archival UVES and ESPRESSO spectra, and simulated two campaigns: monthly monitoring in 2026--2027 and a 25-epoch campaign around periastron. Our main results are the following. 

\begin{enumerate} 

\item Line strengths are primarily determined by $\fEdd$, not $\Mdw$. In the dense models, the emission comes from gas in free fall onto the BH and is spread over thousands of \kms, so that a narrow-line search recovers only about one percent of it. H$\alpha$ is the constraining line; no ultraviolet line is observable below the Chandra bound. 

\item The archival UVES and ESPRESSO spectra exclude only three of the 28 slow-wind models ($\Mdw<2\times10^{-9}$ and $<2.6\times10^{-10}\,\Msunyr$ at $\fEdd=2\times10^{-6}$ and $10^{-5}$, 95\%), all at accretion rates that the X-ray bound already rules out: within the ADAF model they add nothing to it. No fast-wind model is constrained. 

\item Monitoring in 2026--2027 would detect at $5\sigma$ only what the archival data already disfavor. The 2029 periastron campaign significantly improves the constraining power: for a slow wind, a detection would show the accretion state rising two- to threefold along the orbit; for a fast wind, whose capture rate rises 14-fold toward periastron, it detects $\Mdw\simeq10^{-8}\,\Msunyr$ for $\fEdd\gtrsim1.2\times10^{-6}$, a region the apastron X-ray bound allows. With a contemporaneous X-ray $\fEdd$, a detection gives $\Mdw$ and $\epsilon$. 

\item The X-ray bound alone implies $\epsilon<0.2$ for $\Mdw=10^{-11}\,\Msunyr$ and $<2\times10^{-4}$ for $10^{-8}\,\Msunyr$; the lines give $\epsilon<3\times10^{-3}$ if $\Mdw\gtrsim2\times10^{-9}\,\Msunyr$. Since the flow loses most of its gas to an outflow inside $r_{\rm hot}$, it can take in 22 times more than reaches the black hole, so that near the Reimers rate the X-ray bound allows the whole captured wind to enter it. A slow wind accreting below the Chandra bound produces no detectable line; its mass-loss rate has to be measured in ultraviolet absorption against the giant.

\end{enumerate}

In summary, within the ADAF model, the optical spectra we already have add nothing to the X-ray bound, and only the 2029 periastron can, if accretion onto Gaia~BH3 responds to its orbit or if the wind is fast. Gaia~BH3 will not return to periastron until 2041. The 2029 passage is the first, and for a long time the only, opportunity to watch a black hole of known mass respond to a known change in its gas supply. Combined with a measurement of the wind itself in ultraviolet absorption, it could make Gaia~BH3 the first stellar-mass black hole for which we directly measure the fraction of available gas that is actually accreted.

\begin{acknowledgments}
N.C. and F.P. acknowledge support from NASA through Chandra Award No. GO3-24087A. F.P. acknowledges JWST Award No. JWST-GO-03805.010-A. The authors used generative AI under full human supervision: Claude (Anthropic; Claude Opus 5.5) was used to proofread the manuscript text and to polish figures.
\end{acknowledgments}

\facilities{VLT:Kueyen (UVES), VLT (ESPRESSO), CXO}
\software{Cloudy C23.01 \citep{chatzikos23}, LLAGNSED \citep{pesce21}, NumPy, SciPy, Matplotlib}

\bibliographystyle{aasjournalv7}
\bibliography{refs}

@ARTICLE{panuzzo24,
       author = {{Gaia Collaboration} and {Panuzzo}, P. and {Mazeh}, T. and {Arenou}, F. and {Holl}, B. and {Caffau}, E. and {Jorissen}, A. and {Babusiaux}, C. and {Gavras}, P. and {Sahlmann}, J. and {Bastian}, U. and {Wyrzykowski}, {\L}. and {Eyer}, L. and {Leclerc}, N. and {Bauchet}, N. and {Bombrun}, A. and {Mowlavi}, N. and {Seabroke}, G.~M. and {Teyssier}, D. and {Balbinot}, E. and {Helmi}, A. and {Brown}, A.~G.~A. and {Vallenari}, A. and {Prusti}, T. and {de Bruijne}, J.~H.~J. and {Barbier}, A. and {Biermann}, M. and {Creevey}, O.~L. and {Ducourant}, C. and {Evans}, D.~W. and {Guerra}, R. and {Hutton}, A. and {Jordi}, C. and {Klioner}, S.~A. and {Lammers}, U. and {Lindegren}, L. and {Luri}, X. and {Mignard}, F. and {Nicolas}, C. and {Randich}, S. and {Sartoretti}, P. and {Smiljanic}, R. and {Tanga}, P. and {Walton}, N.~A. and {Aerts}, C. and {Bailer-Jones}, C.~A.~L. and {Cropper}, M. and {Drimmel}, R. and {Jansen}, F. and {Katz}, D. and {Lattanzi}, M.~G. and {Soubiran}, C. and {Th{\'e}venin}, F. and {van Leeuwen}, F. and {Andrae}, R. and {Audard}, M. and {Bakker}, J. and {Blomme}, R. and {Casta{\~n}eda}, J. and {De Angeli}, F. and {Fabricius}, C. and {Fouesneau}, M. and {Fr{\'e}mat}, Y. and {Galluccio}, L. and {Guerrier}, A. and {Heiter}, U. and {Masana}, E. and {Messineo}, R. and {Nienartowicz}, K. and {Pailler}, F. and {Riclet}, F. and {Roux}, W. and {Sordo}, R. and {Gracia-Abril}, G. and {Portell}, J. and {Altmann}, M. and {Benson}, K. and {Berthier}, J. and {Burgess}, P.~W. and {Busonero}, D. and {Busso}, G. and {Cacciari}, C. and {C{\'a}novas}, H. and {Carrasco}, J.~M. and {Carry}, B. and {Cellino}, A. and {Cheek}, N. and {Clementini}, G. and {Damerdji}, Y. and {Davidson}, M. and {de Teodoro}, P. and {Delchambre}, L. and {Dell'Oro}, A. and {Fraile Garcia}, E. and {Garabato}, D. and {Garc{\'\i}a-Lario}, P. and {Haigron}, R. and {Hambly}, N.~C. and {Harrison}, D.~L. and {Hatzidimitriou}, D. and {Hern{\'a}ndez}, J. and {Hestroffer}, D. and {Hodgkin}, S.~T. and {Jamal}, S. and {Jevardat de Fombelle}, G. and {Jordan}, S. and {Krone-Martins}, A. and {Lanzafame}, A.~C. and {L{\"o}ffler}, W. and {Lorca}, A. and {Marchal}, O. and {Marrese}, P.~M. and {Moitinho}, A. and {Muinonen}, K. and {Nu{\~n}ez Campos}, M. and {Oreshina-Slezak}, I. and {Osborne}, P. and {Pancino}, E. and {Pauwels}, T. and {Recio-Blanco}, A. and {Riello}, M. and {Rimoldini}, L. and {Robin}, A.~C. and {Roegiers}, T. and {Sarro}, L.~M. and {Schultheis}, M. and {Smith}, M. and {Sozzetti}, A. and {Utrilla}, E. and {van Leeuwen}, M. and {Weingrill}, K. and {Abbas}, U. and {{\'A}brah{\'a}m}, P. and {Abreu Aramburu}, A. and {Ahmed}, S. and {Altavilla}, G. and {{\'A}lvarez}, M.~A. and {Anders}, F. and {Anderson}, R.~I. and {Anglada Varela}, E. and {Antoja}, T. and {Baig}, S. and {Baines}, D. and {Baker}, S.~G. and {Balaguer-N{\'u}{\~n}ez}, L. and {Balog}, Z. and {Barache}, C. and {Barros}, M. and {Barstow}, M.~A. and {Bartolom{\'e}}, S. and {Bashi}, D. and {Bassilana}, J.-L. and {Baudeau}, N. and {Becciani}, U. and {Bedin}, L.~R. and {Bellas-Velidis}, I. and {Bellazzini}, M. and {Beordo}, W. and {Bernet}, M. and {Bertolotto}, C. and {Bertone}, S. and {Bianchi}, L. and {Binnenfeld}, A. and {Blanco-Cuaresma}, S. and {Bland-Hawthorn}, J. and {Blazere}, A. and {Boch}, T. and {Bossini}, D. and {Bouquillon}, S. and {Bragaglia}, A. and {Braine}, J. and {Bratsolis}, E. and {Breedt}, E. and {Bressan}, A. and {Brouillet}, N. and {Brugaletta}, E. and {Bucciarelli}, B. and {Butkevich}, A.~G. and {Buzzi}, R. and {Camut}, A. and {Cancelliere}, R. and {Cantat-Gaudin}, T. and {Capilla Guilarte}, D. and {Carballo}, R. and {Carlucci}, T. and {Carnerero}, M.~I. and {Carretero}, J. and {Carton}, S. and {Casamiquela}, L. and {Casey}, A. and {Castellani}, M. and {Castro-Ginard}, A. and {Ceraj}, L. and {Cesare}, V. and {Charlot}, P. and {Chaudet}, C. and {Chemin}, L. and {Chiavassa}, A. and {Chornay}, N. and {Chosson}, D.},
        title = "{Discovery of a dormant 33 solar-mass black hole in pre-release Gaia astrometry}",
      journal = {\aap},
         year = 2024,
        month = jun,
       volume = {686},
          eid = {L2},
        pages = {L2},
          doi = {10.1051/0004-6361/202449763},
archivePrefix = {arXiv},
       eprint = {2404.10486},
 primaryClass = {astro-ph.GA},
       adsurl = {https://ui.adsabs.harvard.edu/abs/2024A&A...686L...2G}
}

@ARTICLE{pacucci25,
       author = {{Pacucci}, Fabio and {Loeb}, Abraham},
        title = "{Cosmic Outliers: Low-spin Halos Explain the Abundance, Compactness, and Redshift Evolution of the Little Red Dots}",
      journal = {\apjl},
         year = 2025,
        month = aug,
       volume = {989},
       number = {2},
          eid = {L19},
        pages = {L19},
          doi = {10.3847/2041-8213/ade871},
archivePrefix = {arXiv},
       eprint = {2506.03244},
 primaryClass = {astro-ph.GA},
       adsurl = {https://ui.adsabs.harvard.edu/abs/2025ApJ...989L..19P}
}

@ARTICLE{llerena25,
       author = {{Llerena}, M. and {Pentericci}, L. and {Napolitano}, L. and {Mascia}, S. and {Amor{\'\i}n}, R. and {Calabr{\`o}}, A. and {Castellano}, M. and {Cleri}, N.~J. and {Giavalisco}, M. and {Grogin}, N.~A. and {Hathi}, N.~P. and {Hirschmann}, M. and {Koekemoer}, A.~M. and {Nanayakkara}, T. and {Pacucci}, F. and {Shen}, L. and {Wilkins}, S.~M. and {Yoon}, I. and {Yung}, L.~Y.~A. and {Bhatawdekar}, R. and {Lucas}, R.~A. and {Wang}, X. and {Arrabal Haro}, P. and {Bagley}, M.~B. and {Finkelstein}, S.~L. and {Kartaltepe}, J.~S. and {Merlin}, E. and {Papovich}, C. and {Pirzkal}, N. and {Santini}, P.},
        title = "{The ionizing photon production efficiency of star-forming galaxies at z {\ensuremath{\sim}} 4─10}",
      journal = {\aap},
         year = 2025,
        month = jun,
       volume = {698},
          eid = {A302},
        pages = {A302},
          doi = {10.1051/0004-6361/202453251},
archivePrefix = {arXiv},
       eprint = {2412.01358},
 primaryClass = {astro-ph.GA},
       adsurl = {https://ui.adsabs.harvard.edu/abs/2025A&A...698A.302L}
}

@ARTICLE{saladino18,
       author = {{Saladino}, M.~I. and {Pols}, O.~R. and {van der Helm}, E. and {Pelupessy}, I. and {Portegies Zwart}, S.},
        title = "{Gone with the wind: the impact of wind mass transfer on the orbital evolution of AGB binary systems}",
      journal = {\aap},
         year = 2018,
        month = oct,
       volume = {618},
          eid = {A50},
        pages = {A50},
          doi = {10.1051/0004-6361/201832967},
archivePrefix = {arXiv},
       eprint = {1805.03208},
 primaryClass = {astro-ph.SR},
       adsurl = {https://ui.adsabs.harvard.edu/abs/2018A&A...618A..50S}
}

@ARTICLE{abate13,
       author = {{Abate}, C. and {Pols}, O.~R. and {Izzard}, R.~G. and {Mohamed}, S.~S. and {de Mink}, S.~E.},
        title = "{Wind Roche-lobe overflow: Application to carbon-enhanced metal-poor stars}",
      journal = {\aap},
         year = 2013,
        month = apr,
       volume = {552},
          eid = {A26},
        pages = {A26},
          doi = {10.1051/0004-6361/201220007},
archivePrefix = {arXiv},
       eprint = {1302.4441},
 primaryClass = {astro-ph.SR},
       adsurl = {https://ui.adsabs.harvard.edu/abs/2013A&A...552A..26A}
}

@INPROCEEDINGS{mohamed07,
       author = {{Mohamed}, S. and {Podsiadlowski}, Ph.},
        title = "{Wind Roche-Lobe Overflow: a New Mass-Transfer Mode for Wide Binaries}",
    booktitle = {15th European Workshop on White Dwarfs},
         year = 2007,
       editor = {{Napiwotzki}, R. and {Burleigh}, M.~R.},
       series = {Astronomical Society of the Pacific Conference Series},
       volume = {372},
        month = sep,
        pages = {397},
       adsurl = {https://ui.adsabs.harvard.edu/abs/2007ASPC..372..397M}
}

@ARTICLE{dupree09,
       author = {{Dupree}, A.~K. and {Smith}, Graeme H. and {Strader}, Jay},
        title = "{Fast Winds and Mass Loss from Metal-Poor Field Giants}",
      journal = {\aj},
         year = 2009,
        month = nov,
       volume = {138},
       number = {5},
        pages = {1485-1501},
          doi = {10.1088/0004-6256/138/5/1485},
archivePrefix = {arXiv},
       eprint = {0909.1558},
 primaryClass = {astro-ph.SR},
       adsurl = {https://ui.adsabs.harvard.edu/abs/2009AJ....138.1485D}
}

@ARTICLE{Elbert_2018,
       author = {{Elbert}, Oliver D. and {Bullock}, James S. and {Kaplinghat}, Manoj},
        title = "{Counting black holes: The cosmic stellar remnant population and implications for LIGO}",
      journal = {\mnras},
         year = 2018,
        month = jan,
       volume = {473},
       number = {1},
        pages = {1186-1194},
          doi = {10.1093/mnras/stx1959},
archivePrefix = {arXiv},
       eprint = {1703.02551},
 primaryClass = {astro-ph.GA},
       adsurl = {https://ui.adsabs.harvard.edu/abs/2018MNRAS.473.1186E}
}

@ARTICLE{cappelluti24,
       author = {{Cappelluti}, Nico and {Pacucci}, Fabio and {Hasinger}, G{\"u}nther},
        title = "{Constraining Wind-driven Accretion onto Gaia BH3 with Chandra}",
      journal = {\apj},
         year = 2024,
        month = oct,
       volume = {973},
       number = {2},
          eid = {75},
        pages = {75},
          doi = {10.3847/1538-4357/ad6f96},
archivePrefix = {arXiv},
       eprint = {2406.07602},
 primaryClass = {astro-ph.HE},
       adsurl = {https://ui.adsabs.harvard.edu/abs/2024ApJ...973...75C}
}

@ARTICLE{pesce21,
       author = {{Pesce}, Dominic W. and {Palumbo}, Daniel C.~M. and {Narayan}, Ramesh and {Blackburn}, Lindy and
                 {Doeleman}, Sheperd S. and {Johnson}, Michael D. and {Ma}, Chung-Pei and {Nagar}, Neil M. and
                 {Natarajan}, Priyamvada and {Ricarte}, Angelo},
        title = "{Toward Determining the Number of Observable Supermassive Black Hole Shadows}",
      journal = {\apj},
         year = 2021,
        month = dec,
       volume = {923},
       number = {2},
          eid = {260},
        pages = {260},
          doi = {10.3847/1538-4357/ac2eb5},
archivePrefix = {arXiv},
       eprint = {2108.05228},
 primaryClass = {astro-ph.HE},
       adsurl = {https://ui.adsabs.harvard.edu/abs/2021ApJ...923..260P}
}

@ARTICLE{mahadevan97,
       author = {{Mahadevan}, Rohan},
        title = "{Scaling Laws for Advection-dominated Flows: Applications to Low-Luminosity Galactic Nuclei}",
      journal = {\apj},
         year = 1997,
        month = mar,
       volume = {477},
       number = {2},
        pages = {585-601},
          doi = {10.1086/303727},
archivePrefix = {arXiv},
       eprint = {astro-ph/9609107},
 primaryClass = {astro-ph},
       adsurl = {https://ui.adsabs.harvard.edu/abs/1997ApJ...477..585M}
}

@ARTICLE{narayan95,
       author = {{Narayan}, Ramesh and {Yi}, Insu},
        title = "{Advection-dominated Accretion: Underfed Black Holes and Neutron Stars}",
      journal = {\apj},
         year = 1995,
        month = oct,
       volume = {452},
        pages = {710},
          doi = {10.1086/176343},
archivePrefix = {arXiv},
       eprint = {astro-ph/9411059},
 primaryClass = {astro-ph},
       adsurl = {https://ui.adsabs.harvard.edu/abs/1995ApJ...452..710N}
}

@ARTICLE{shakura73,
       author = {{Shakura}, N.~I. and {Sunyaev}, R.~A.},
        title = "{Black holes in binary systems. Observational appearance.}",
      journal = {\aap},
         year = 1973,
        month = jan,
       volume = {24},
        pages = {337-355},
       adsurl = {https://ui.adsabs.harvard.edu/abs/1973A&A....24..337S}
}

@ARTICLE{chatzikos23,
       author = {{Chatzikos}, M. and {Bianchi}, S. and {Camilloni}, F. and {Chakraborty}, P. and {Gunasekera}, C.~M. and
                 {Guzm{\'a}n}, F. and {Milby}, J.~S. and {Sarkar}, A. and {Shaw}, G. and {van Hoof}, P.~A.~M. and
                 {Ferland}, G.~J.},
        title = "{The 2023 Release of Cloudy}",
      journal = {\rmxaa},
         year = 2023,
        month = oct,
       volume = {59},
        pages = {327-343},
          doi = {10.22201/ia.01851101p.2023.59.02.12},
archivePrefix = {arXiv},
       eprint = {2308.06396},
 primaryClass = {astro-ph.GA},
       adsurl = {https://ui.adsabs.harvard.edu/abs/2023RMxAA..59..327C}
}

@ARTICLE{grevesse10,
       author = {{Grevesse}, N. and {Asplund}, M. and {Sauval}, A.~J. and {Scott}, P.},
        title = "{The chemical composition of the Sun}",
      journal = {\apss},
         year = 2010,
        month = jul,
       volume = {328},
       number = {1-2},
        pages = {179-183},
          doi = {10.1007/s10509-010-0288-z},
       adsurl = {https://ui.adsabs.harvard.edu/abs/2010Ap&SS.328..179G}
}

@ARTICLE{hatchett77,
       author = {{Hatchett}, S. and {McCray}, R.},
        title = "{X-ray sources in stellar winds.}",
      journal = {\apj},
         year = 1977,
        month = jan,
       volume = {211},
        pages = {552-561},
          doi = {10.1086/154962},
       adsurl = {https://ui.adsabs.harvard.edu/abs/1977ApJ...211..552H}
}

@ARTICLE{reimers75,
       author = {{Reimers}, D.},
        title = "{Circumstellar absorption lines and mass loss from red giants.}",
      journal = {Memoires of the Societe Royale des Sciences de Liege},
         year = 1975,
        month = jan,
       volume = {8},
        pages = {369-382},
       adsurl = {https://ui.adsabs.harvard.edu/abs/1975MSRSL...8..369R}
}

@ARTICLE{castor79,
       author = {{Castor}, J.~I. and {Lamers}, H.~J.~G.~L.~M.},
        title = "{An atlas of theoretical P Cygni profiles.}",
      journal = {\apjs},
         year = 1979,
        month = apr,
       volume = {39},
        pages = {481-511},
          doi = {10.1086/190583},
       adsurl = {https://ui.adsabs.harvard.edu/abs/1979ApJS...39..481C}
}

@BOOK{lamers99,
       author = {{Lamers}, Henny J.~G.~L.~M. and {Cassinelli}, Joseph P.},
        title = "{Introduction to Stellar Winds}",
    publisher = {Cambridge University Press},
      address = {Cambridge},
         year = 1999,
       adsurl = {https://ui.adsabs.harvard.edu/abs/1999isw..book.....L}
}

@BOOK{osterbrock06,
       author = {{Osterbrock}, Donald E. and {Ferland}, Gary J.},
        title = "{Astrophysics of gaseous nebulae and active galactic nuclei}",
    publisher = {University Science Books},
      address = {Sausalito, CA},
         year = 2006,
       adsurl = {https://ui.adsabs.harvard.edu/abs/2006agna.book.....O}
}

@ARTICLE{voronov97,
       author = {{Voronov}, G.~S.},
        title = "{A Practical Fit Formula for Ionization Rate Coefficients of Atoms and Ions by Electron Impact: Z = 1-28}",
      journal = {Atomic Data and Nuclear Data Tables},
         year = 1997,
        month = jan,
       volume = {65},
        pages = {1},
          doi = {10.1006/adnd.1997.0732},
       adsurl = {https://ui.adsabs.harvard.edu/abs/1997ADNDT..65....1V}
}

@ARTICLE{hui97,
       author = {{Hui}, Lam and {Gnedin}, Nickolay Y.},
        title = "{Equation of state of the photoionized intergalactic medium}",
      journal = {\mnras},
         year = 1997,
        month = nov,
       volume = {292},
       number = {1},
        pages = {27-42},
          doi = {10.1093/mnras/292.1.27},
archivePrefix = {arXiv},
       eprint = {astro-ph/9612232},
 primaryClass = {astro-ph},
       adsurl = {https://ui.adsabs.harvard.edu/abs/1997MNRAS.292...27H}
}

@ARTICLE{cen92,
       author = {{Cen}, Renyue},
        title = "{A Hydrodynamic Approach to Cosmology: Methodology}",
      journal = {\apjs},
         year = 1992,
        month = feb,
       volume = {78},
        pages = {341},
          doi = {10.1086/191630},
       adsurl = {https://ui.adsabs.harvard.edu/abs/1992ApJS...78..341C}
}

@ARTICLE{shull85,
       author = {{Shull}, J.~M. and {van Steenberg}, M.~E.},
        title = "{X-ray secondary heating and ionization in quasar emission-line clouds}",
      journal = {\apj},
         year = 1985,
        month = nov,
       volume = {298},
        pages = {268-274},
          doi = {10.1086/163605},
       adsurl = {https://ui.adsabs.harvard.edu/abs/1985ApJ...298..268S}
}

@ARTICLE{verner96,
       author = {{Verner}, D.~A. and {Ferland}, G.~J. and {Korista}, K.~T. and {Yakovlev}, D.~G.},
        title = "{Atomic Data for Astrophysics. II. New Analytic Fits for Photoionization Cross Sections of Atoms and Ions}",
      journal = {\apj},
         year = 1996,
        month = jul,
       volume = {465},
        pages = {487},
          doi = {10.1086/177435},
archivePrefix = {arXiv},
       eprint = {astro-ph/9601009},
 primaryClass = {astro-ph},
       adsurl = {https://ui.adsabs.harvard.edu/abs/1996ApJ...465..487V}
}

@BOOK{draine11,
       author = {{Draine}, Bruce T.},
        title = "{Physics of the Interstellar and Intergalactic Medium}",
    publisher = {Princeton University Press},
      address = {Princeton, NJ},
         year = 2011,
       adsurl = {https://ui.adsabs.harvard.edu/abs/2011piim.book.....D}
}

@ARTICLE{gallo06,
       author = {{Gallo}, E. and {Fender}, R.~P. and {Miller-Jones}, J.~C.~A. and {Merloni}, A. and {Jonker}, P.~G. and {Heinz}, S. and
                 {Maccarone}, T.~J. and {van der Klis}, M.},
        title = "{A radio-emitting outflow in the quiescent state of A0620-00: implications for modelling low-luminosity black hole binaries}",
      journal = {\mnras},
         year = 2006,
        month = aug,
       volume = {370},
       number = {3},
        pages = {1351-1360},
          doi = {10.1111/j.1365-2966.2006.10560.x},
archivePrefix = {arXiv},
       eprint = {astro-ph/0605376},
 primaryClass = {astro-ph},
       adsurl = {https://ui.adsabs.harvard.edu/abs/2006MNRAS.370.1351G}
}

@ARTICLE{yuan14,
       author = {{Yuan}, Feng and {Narayan}, Ramesh},
        title = "{Hot Accretion Flows Around Black Holes}",
      journal = {\araa},
         year = 2014,
        month = aug,
       volume = {52},
        pages = {529-588},
          doi = {10.1146/annurev-astro-082812-141003},
archivePrefix = {arXiv},
       eprint = {1401.0586},
 primaryClass = {astro-ph.HE},
       adsurl = {https://ui.adsabs.harvard.edu/abs/2014ARA&A..52..529Y}
}

@ARTICLE{olejak20,
       author = {{Olejak}, A. and {Belczynski}, K. and {Bulik}, T. and {Sobolewska}, M.},
        title = "{Synthetic catalog of black holes in the Milky Way}",
      journal = {\aap},
         year = 2020,
        month = jun,
       volume = {638},
          eid = {A94},
        pages = {A94},
          doi = {10.1051/0004-6361/201936557},
archivePrefix = {arXiv},
       eprint = {1908.08775},
 primaryClass = {astro-ph.SR},
       adsurl = {https://ui.adsabs.harvard.edu/abs/2020A&A...638A..94O}
}

@ARTICLE{elbadry23a,
       author = {{El-Badry}, Kareem and {Rix}, Hans-Walter and {Quataert}, Eliot and {Howard}, Andrew W. and {Isaacson}, Howard and
                 {Fuller}, Jim and {Hawkins}, Keith and {Breivik}, Katelyn and {Wong}, Kaze W.~K. and {Rodriguez}, Antonio C. and others},
        title = "{A Sun-like star orbiting a black hole}",
      journal = {\mnras},
         year = 2023,
        month = jan,
       volume = {518},
       number = {1},
        pages = {1057-1085},
          doi = {10.1093/mnras/stac3140},
archivePrefix = {arXiv},
       eprint = {2209.06833},
 primaryClass = {astro-ph.SR},
       adsurl = {https://ui.adsabs.harvard.edu/abs/2023MNRAS.518.1057E}
}

@ARTICLE{elbadry23b,
       author = {{El-Badry}, Kareem and {Rix}, Hans-Walter and {Cendes}, Yvette and {Rodriguez}, Antonio C. and {Conroy}, Charlie and
                 {Quataert}, Eliot and {Hawkins}, Keith and {Zari}, Eleonora and {Hobson}, Melissa and {Breivik}, Katelyn and others},
        title = "{A red giant orbiting a black hole}",
      journal = {\mnras},
         year = 2023,
        month = may,
       volume = {521},
       number = {3},
        pages = {4323-4348},
          doi = {10.1093/mnras/stad799},
archivePrefix = {arXiv},
       eprint = {2302.07880},
 primaryClass = {astro-ph.SR},
       adsurl = {https://ui.adsabs.harvard.edu/abs/2023MNRAS.521.4323E}
}

@ARTICLE{agol02,
       author = {{Agol}, Eric and {Kamionkowski}, Marc},
        title = "{X-rays from isolated black holes in the Milky Way}",
      journal = {\mnras},
         year = 2002,
        month = aug,
       volume = {334},
       number = {3},
        pages = {553-562},
          doi = {10.1046/j.1365-8711.2002.05523.x},
archivePrefix = {arXiv},
       eprint = {astro-ph/0109539},
 primaryClass = {astro-ph},
       adsurl = {https://ui.adsabs.harvard.edu/abs/2002MNRAS.334..553A}
}

@ARTICLE{mcdonald15,
       author = {{McDonald}, I. and {Zijlstra}, A.~A.},
        title = "{Mass-loss on the red giant branch: the value and metallicity dependence of Reimers' {\ensuremath{\eta}} in globular clusters}",
      journal = {\mnras},
         year = 2015,
        month = mar,
       volume = {448},
       number = {1},
        pages = {502-521},
          doi = {10.1093/mnras/stv007},
archivePrefix = {arXiv},
       eprint = {1501.00874},
 primaryClass = {astro-ph.SR},
       adsurl = {https://ui.adsabs.harvard.edu/abs/2015MNRAS.448..502M}
}

@ARTICLE{merloni03,
       author = {{Merloni}, Andrea and {Heinz}, Sebastian and {di Matteo}, Tiziana},
        title = "{A Fundamental Plane of black hole activity}",
      journal = {\mnras},
         year = 2003,
        month = nov,
       volume = {345},
       number = {4},
        pages = {1057-1076},
          doi = {10.1046/j.1365-2966.2003.07017.x},
archivePrefix = {arXiv},
       eprint = {astro-ph/0305261},
 primaryClass = {astro-ph},
       adsurl = {https://ui.adsabs.harvard.edu/abs/2003MNRAS.345.1057M}
}

@ARTICLE{sjouwerman24,
       author = {{Sjouwerman}, L.~O. and {Blanchard}, J.~M.},
        title = "{Tighter upper limits to radio emission from GAIA BH3 with the VLA}",
      journal = {The Astronomer's Telegram},
         year = 2024,
        month = sep,
       volume = {16832},
        pages = {1},
       adsurl = {https://ui.adsabs.harvard.edu/abs/2024ATel16832....1S}
}

\end{document}